\documentclass[twocolumn]{aastex631}
\usepackage{bm}
\usepackage{amsmath}	% Advanced maths commandshttps://www.overleaf.com/project/5f3311dbac706300018f069a
\usepackage{amssymb}	% Extra maths symbols

\providecommand{\abs}[1]{\lvert#1\rvert}

\shorttitle{On wave interference in planet migration}
\shortauthors{Chametla et al.}
\graphicspath{{./}{figures/}}
\begin{document}

%\title{Template \aastex Article with Examples: 
%v6.3.1\footnote{Released on March, 1st, 2021}}

\title{On wave interference in planet migration: dead zone torques modified by active zone forcing}

\correspondingauthor{Ra\'{u}l O. Chametla}
\email{raul@sirrah.troja.mff.cuni.cz}

\author[0000-0002-4327-7857]{Raúl O. Chametla}
\affiliation{Charles University, Faculty of Mathematics and Physics, Astronomical Institute \\ 
V Hole$\check{s}$ovi$\check{c}$k\'ach 747/2, 180 00 \\ 
Prague 8, Czech Republic}
\author[0000-0001-7215-5026]{Ondrej Chrenko}
\affiliation{Charles University, Faculty of Mathematics and Physics, Astronomical Institute \\ 
V Hole$\check{s}$ovi$\check{c}$k\'ach 747/2, 180 00 \\ 
Prague 8, Czech Republic}

\author[0000-0002-3768-7542]{Wladimir Lyra}
\affiliation{Department of Astronomy, New Mexico State University, PO Box 30001, MSC 4500 Las Cruces, NM 88003, USA}

\author[0000-0001-8292-1943]{Neal J. Turner}
\affiliation{Jet Propulsion Laboratory, California Institute of Technology, 4800 Oak Grove Drive, Pasadena, California 91109, U.S.A}

%\affiliation{AAS Director of Publishing}
%\affiliation{American Astronomical Society \\
%1667 K Street NW, Suite 800 \\

%\author{Magaret Donnelly}
%\affiliation{IOP Publishing, Washington, DC 20005}



\begin{abstract}
We investigate planetary migration in the dead zone of a protoplanetary disk where there are a set of spiral waves propagating inward due to the turbulence in the active zone and the Rossby wave instability (RWI), which occurs at the transition between the dead and active zones. We perform global 3D unstratified magnetohydrodynamical (MHD) simulations of a gaseous disk with the FARGO3D code, using weak gradients in the static resistivity profiles that trigger the formation of a vortex at the outer edge of the dead zone. We find that once the Rossby vortex develops, spiral waves in the dead zone emerge and interact with embedded migrating planets by wave interference, which notably changes their migration. The inward migration becomes faster depending on the mass of the planet, due mostly to the constructive (destructive) interference between the outer (inner) spiral arm of the planet and, the destruction of the dynamics of the horseshoe region by means of the set of background spiral waves propagating inward. The constructive wave interference produces a more negative Lindblad differential torque which inevitably leads to an inward migration. Lastly, for massive planets embedded in the dead zone, we find that the spiral waves can create an asymmetric wider and depeer gap than in the case of $\alpha$-disks, and can prevent the formation of vortices at the outer edge of the gap. The latter could generate a faster or slower migration compared to the standard type-II migration.
\end{abstract}


\section{Introduction} \label{sec:intro}
Early studies about the gravitational interaction between planets and protoplanetary disks including the Magneto-Rotational Instability \citep[MRI;][]{BH1991} were presented by \citet{PN2003}, \citet[][]{NP2003,NP2004,PNS2004}. For low-mass planets embedded in turbulent disks \citet{NP2004} using three dimensional (3D) global and local shearing box ideal magnetohydrodynamical (MHD) simulations, found that net torque experienced by a protoplanet shows strong fluctuations oscillating between negative and positive values within an orbital timescale, suggesting that these planets would experience a random walk behavior in their migration. 

In a different approach, \citet{LSA2004} performed 2D hydrodynamic (HD) simulations (parameterized via 3D-MHD simulations) with forced turbulence excited through a stirring potential. They found similar conclusions about the random walk migration.

\citet{Nelson2005} studied the orbital evolution of “planetesimals" and planets with masses $M_p\in[0,30]M_\oplus$ (here $M_\oplus$ is the Earth mass) embedded in a turbulent disk and found that over the entire range of masses considered, migration is stochastic due to gravitational interaction with turbulent density fluctuations in the disk. In addition, he found that the eccentricities are excited by the turbulence, from 0.02 up to 0.14 depending on the mass of the planet. \citet{NP2003} and \citet{Zhu_etal2013} found that the gap carved in a turbulent disk by a massive planet is deeper and wider in comparison with laminar disk models when azimuthal and vertical magnetic fields are considered. In the end, similar conclusions are reached in all these studies that the MHD turbulence will alter the migration timescales in comparison with the standard types of planetary migration \citep[see][for a review]{KN2012,BM2013,Lesur_etal2022,Paardekooper_etal2022}. 

On the other hand, recent studies have focused on the possible effects induced in the dead zone from turbulence in the active zone of the disk. Motivated by the previous results of the random-walk migration, \citet{OMM2007} performed 3D stratified local shearing box  MHD simulations with magnetorotational turbulence confined to thin active layers above and below the midplane. They found that the gravitational torques on low-mass planets located in the midplane exhibits turbulent fluctuations which are due to the turbulence in the active layers of the protoplanetary disk. 

In addition, radial gas flows can occur in the dead zone due to non-linear MHD effects \citep{,Bai2014ApJ,Lesur_etal2014,XB2016,Bethune_etal2017}. The flow-induced asymmetric distortion of the corotation region creates a dynamical corotation torque \citep[][]{McNally_etal2017,MNP2018,McNally_etal2018}. This dynamical corotation torque can induce up to four different migration regimes for low-mass planets (parameterized by the relative radial velocity between the gas and the planet).

Focused on this direction, \citet{McNally_etal2020} performed 3D inviscid simulations with and without laminar accretion flows, considering the thermodynamics of the disk very close to the adiabatic case \citep[that is, employed a temperature forcing like that used by][]{Lyra_etal2016}. They found that a wind-driven laminar accretion flow through the surface layers of the disk does not significantly modify the total torque.
However, the dynamical corotation torque produces a faster inward migration contrary to the results obtained previously in 2D simulations when the radial flow of gas is directed towards the star.

Another interesting effect that can occur in the dead zone is the formation of a spiral pattern which is reminiscent of planet-induced spiral arms \citep{Lyra_etal2015}. This pattern is a result of waves propagating inwards from the turbulent perturbations in the active zone and also from the Rossby vortex formation at the outer dead zone transition. The latter is a consequence of the Rossby Wave Instability \citep[RWI;][]{Lovelace_etal1999,Li_etal2000,Li_etal2001} which arises from radial gradients in the hydrodynamical quantities (for instance, the pressure and the gas density) which are induced through radial transport of the gas by an accretion stress that has a step in the transition regions. This step in the accretion stress can come from a weak gradient in the Ohmic resistivity \citep{Dzyurkevich_etal2010,Lyra_Mac2012,Lyra_etal2015}.

Direct planet-vortex interaction can considerably modify the migration of planets \citep{Regaly_etal2013,Ataiee_etal2014,Faure_etal2016,McNally_etal2019}. For instance, \citet{McNally_etal2019} found that the migration of partial gap-opening planets becomes chaotic for sufficiently low viscosities. They drew a conceptual “city map" of planetary migration mechanisms in 2D-isothermal disks in the viscosity-planet mass physical parameter subspace, and showed that their results can be included in different places on this “city map".

However, the planet-vortex interaction can be also indirect, as demonstrated in \citet{Chametla_Chrenko2022} who studied the impact of vortices formed after destabilization of two pressure bumps on planetary migration. They found that the vortex-induced spiral waves facilitate much slower and/or stagnant migration of the planets and, in some cases, excitation of planetary eccentricities. They called this mechanism the Faraway Interaction (FI). Therefore, considering the previous results of \citet{Lyra_etal2015}, the objective of this work is to study the effect of the background spiral waves on planetary migration in the dead zone of the protoplanetary disk where the formation of vortices and spiral arms is due to a smooth resistivity transition, which is a physically self-consistent scenario.

%%%%% Table 1 %%%%%<<<<<
\begin{table}
    \centering
    \caption{Initial conditions and main parameters of our simulations ($r_0$ denotes a reference radius$^*$).}
    \begin{tabular}{lll}
    \hline
    \hline
        & Parameter & Value [code units]\\
        &           &  \\
    \hline
	&  Central star  &  \\
	&                &  \\
	\hline
         Mass of the star & $M_\star$ & 1.0 \\
    \hline
        & Disk &    \\
        &      &     \\
    \hline
        Aspect ratio at $r_0$ & $h$ & 0.1  \\
           Density at $r_0$ & $\rho_0$ & $6\times10^{-4}$ \\
        Surface density slope & $q_{\rho}$ & 1.5  \\
        Temperature slope & $q_T$ & 1.0  \\
        Initial resistivity & $\eta_0$ &$\{2.28,5\}\times10^{-4}$ \\
        Initial magnetic field$^{**}$ & $B_0$ & $\{4.47\times10^{-3},10^{-2}\}$  \\
        Magnetic permitivity \\of vacuum& $\mu_0$ & 1.0  \\
        Plasma parameter at \\$r_0$ & $\beta$ &$\{2\times10^{2},10^{3}\}$  \\
        Kinematic viscosity \\(only 2D $\alpha$-disks) & $\nu_0$ & $2.56\times10^{-6}$  \\
    \hline
		& Planet  &\\
		&         &\\
	\hline
         Planet mass interval & $M_p$ & $[3\times10^{-6},10^{-2}]$  \\
         Planet location & $(r_p,\phi_p,z_p)$ & $(1,0,0)$ \\
         Softening length & $\epsilon$  & $\{0.2,0.6\}H(r_p)$\\
    \hline
		& Global mesh &\\
		&     & \\
	\hline
	     Radial extension & $r$ &$[0.4,9.6]$\\
	     Azimuthal extension & $\phi$ & $[0,2\pi]$ \\
	     Vertical extension & $z$ & [-0.1,0.1] \\
         Radial resolution & $N_r$ & 518 cells \\
         Azimuthal resolution & $N_\phi$ & 1024 cells \\
         Vertical resolution & $N_z$ & 64 cells \\
         Radial spacing & $\log$ & Logarithmic\\
         \hline
		& Additional \\&parameters  &\\
	\hline
	     Reference frame  & C & Corotating \\
	     Frame angular speed & $\Omega(r_p)$ & $\Omega_p$\\
	     Gravitational constant  & G & 1.0 \\
	     Orbital period at $r_p$& $T_0$ & $2\pi\Omega^{-1}_p$\\
    \hline  
    \end{tabular}
    %\caption{Models}
    \tablecomments{$^*$We consider $r_0=10$ au and $M_\star=1M_\odot$ when scaling back to physical units\\$^{**}$We use a net vertical magnetic field}
    \label{tab:condinit}
\end{table}

%%%%%%%%%%%%%%%%%%%%%%%%%<<<<<

%%%--------------- Table 2
%Summary of Numerical Models in Section 
\begin{table*}
	\centering
	\caption{Summary of Numerical Models in Section \ref{sec:results}.}
	\label{tab:models}
	\begin{tabular}{cccccc}
		\hline
		\hline
		Disk model & Planet Mass & Orbit & Density slope & Magnetic field & Viscosity\\
		         & $M_p[M_\oplus,M_{\mathrm{Jup}}^*]$& &$q_\rho$ &plasma$-\beta$ & $\alpha$-viscosity\\
		\hline
		$\alpha$-disk  \\ 
		\hline
		AR1  & $1M_\oplus$ &Migrating & 1.5 & No & $2.5\times10^{-4}$ \\
		AR2 & $10M_\oplus$&Migrating &1.5& No &$2.5\times10^{-4}$\\
		AR2a  & $10M_\oplus$&Migrating &0.5& No  &$2.5\times10^{-4}$\\
		AR3  & $30M_\oplus$&Migrating &1.5& No & $2.5\times10^{-4}$\\
		AR4  & $100M_\oplus$&Migrating &1.5& No & $2.5\times10^{-4}$\\
		AR5  & $1.05M_{\mathrm{Jup}}$&Fixed &1.5& No  & $2.5\times10^{-4}$\\
		AR6  & $3.15M_{\mathrm{Jup}}$&Fixed &1.5& No  & $2.5\times10^{-4}$\\
		AR7  & $10.5M_{\mathrm{Jup}}$&Fixed &1.5& No  & $2.5\times10^{-4}$\\
		\hline
		MHD-disk \\
		\hline  
		MR1  & $1M_\oplus$&Migrating &1.5&$10^{3}$& No\\
		MR2  &  $10M_\oplus$&Migrating&1.5&$10^{3}$&  No\\
		MR2a  &  $10M_\oplus$&Migrating&0.5&$10^{3}$& No\\
		MR2b  &  $10M_\oplus$&Migrating&1.5&$2\times10^{2}$& No\\
		MR3  & $30M_\oplus$&Migrating&1.5&$10^{3}$& No\\
		MR4  & $100M_\oplus$&Migrating &1.5&$10^{3}$& No\\
		MR5  & $1.05M_{\mathrm{Jup}}$&Fixed &1.5&$10^{3}$& No\\
		MR6  & $3.15M_{\mathrm{Jup}}$&Fixed &1.5&$10^{3}$& No\\
		MR7  & $10.5M_{\mathrm{Jup}}$&Fixed &1.5&$10^{3}$& No\\
		\hline
		
	\end{tabular}
	\tablecomments{$^*$Here $M_\mathrm{Jup}\equiv318M_\oplus$ Earth masses.}
		\label{size}
\end{table*}
%%%%%%%%%%-----------------

The paper is organized as follows. In Section \ref{sec:physical}, we present the physical model, code and numerical setup used in our 3D-MHD simulations. In Section \ref{sec:results}, we present the results of our numerical models. We present a brief discussion in Section \ref{sec:discussion}. Concluding remarks can be found in Section \ref{sec:conclusions}.

\section{Physical Model} 
 \label{sec:physical}
We have carried out 3D global unstratified MHD simulations of a gaseous disk harboring a planet inside of the dead zone. Apart from the numerical scheme used in our code (which we discuss below), the physical model and numerical setup used in this study basically follow those in \citet{Lyra_etal2015}, which we briefly recall for convenience. 

%%%%%%%%%%%%%%---------
%\begin{figure}[ht!]
\begin{figure}
\includegraphics[width=0.5\textwidth]{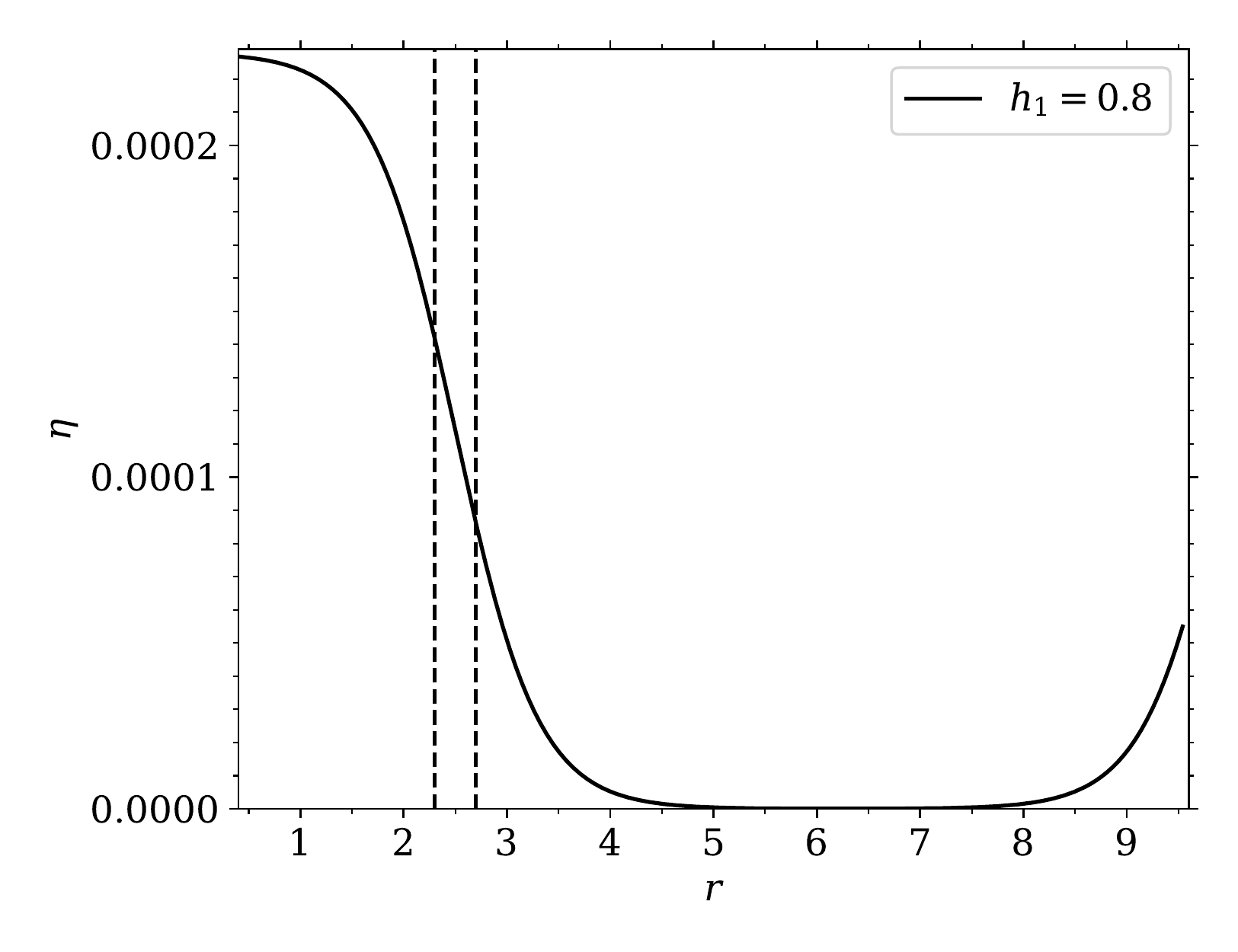}
\caption{Resistivity profile given by Eq. (\ref{eq:eta_pro}) considering $\eta_0=2.28\times10^{-4}$ and a transition width of $h_1=0.8$ centered at $r_1=2.5$. The $\eta$ profile has a smooth jump from $r=1$ up to $r=4$ and corresponds to the smoothest resistivity transition studied in \citet{Lyra_etal2015}. The dashed vertical lines correspond to 2H around of $r_1$.}
\label{fig:eta_dens}
\end{figure}
%%%%%%%%%%%%%%---------

\subsection{Gas disk}

Magnetohydrodynamical equations describing the gas flow in 3D-MHD disks, considering a frame rotating with a uniform angular velocity $\mathbf{\Omega}$, are given by the equation of continuity
\begin{equation}
\frac{\partial\rho}{\partial t}=-(\mathbf{u}\cdot\mathbf{\nabla})\rho-\rho\mathbf{\nabla}\cdot\mathbf{u} \,,
\label{eq:continuity}
\end{equation}
the gas momentum equation
\begin{equation}
\begin{split}
\frac{\partial\mathbf{u}}{\partial t}=-(\mathbf{u}\cdot\mathbf{\nabla})\mathbf{u}-\frac{1}{\rho}\mathbf{\nabla} p-\mathbf{\nabla}\Phi+\frac{\mathbf{J}\times\mathbf{B}}{\rho}
&\\ -\mathbf{\Omega}\times(\mathbf{\Omega}\times\mathbf{r})-2\mathbf{\Omega}\times\mathbf{u}\,,
\label{eq:momentum}
\end{split}
\end{equation}
and the induction equation
\begin{equation}
\frac{\partial \mathbf{B}}{\partial t}=\nabla\times\left[\mathbf{u}\times\mathbf{B}-\eta\mathbf{J} \right],
\label{eq:induction}
\end{equation}
where
\begin{equation}
\rho=\rho_0\left(\frac{r}{r_0}\right)^{-q_\rho}
\label{eq:rho}
\end{equation}
is the gas density, which is related to the surface density in our cylindrical disk approximation by
\begin{equation}
\frac{\Sigma}{L_z}=\frac{1}{2\pi L_z}\int\rho d\phi dz, 
\label{eq:surfdens}
\end{equation}
here $L_z$ represents the vertical extent of our computational domain.

In equations (\ref{eq:continuity}-\ref{eq:momentum}), $\mathbf{u}$ is the gas velocity, $\Phi$ is the gravitational potential, and $p$ is the pressure given as
\begin{equation}
p=\rho c_s^2 \,,
\label{eq:pressure}
\end{equation}
with $c_s$ the sound speed. Note that the word aspect ratio (see Table \ref{tab:condinit}) in unstratified 3D-MHD disks merely implies that $c_s=0.1r\Omega_\mathrm{Kep}$ everywhere in the disk, so disk scale height $H\equiv0.1r$. In the induction equation $\mathbf{J}=\mu_0^{-1}\mathbf{\nabla}\times\mathbf{B}$, is the current density and $\eta$ is the resistivity modeled as in \citet{Lyra_etal2015}
\begin{equation}
\eta(r)=\eta_0-\frac{\eta_0}{2}\left[\tanh\left(\frac{r-r_1}{h_1}\right)-\tanh\left(\frac{r-r_2}{h_2}\right)\right] \, ,
\label{eq:eta_pro}
\end{equation}
which leads to a smooth resistivity transition between the inner dead zone
and the outer active zone.

\subsection{Planet and stellar potentials}
\label{sec:planet}

We use a reference frame centred on the star and rotating with the angular frequency $\mathbf{\Omega}=\Omega_p\mathbf{\hat{k}}$, where $\Omega_p\equiv\sqrt{GM_{\star}/r_p^3}$ is the angular velocity of the planet (with $r_p$ the planet's orbital radius) and $\mathbf{\hat{k}}$ is a unit vector along the rotation axis. The unit of time adopted when discussing the results is $T_0=2\pi\Omega_p^{-1}$. The gravitational potential $\Phi$ is given by
\begin{equation}
\Phi=\Phi_S+\Phi_p+\Phi_{\mathrm{ind}},
 \label{eq:potential}
\end{equation}
where
\begin{equation}
\Phi_S=-\frac{GM_\star}{r},
 \label{eq:Star_potential}
\end{equation}
is the stellar potential with $M_\star$ the mass of the central star and $G$ the gravitational constant,
\begin{equation}
\Phi_p=-\frac{GM_p}{\sqrt{r'^2+\epsilon^2}} \, ,
 \label{eq:Planet_potential}
\end{equation}
is the planetary potential and, 
\begin{equation}
\Phi_{\mathrm{ind}}=\frac{GM_p}{r_p^3}\textbf{r}\cdot\textbf{r}_p+G\int_V\frac{dm(\mathbf{r'})}{r'^3}\textbf{r}\cdot\textbf{r}' \, ,
\label{eq:ind_pot}
\end{equation}
is the indirect potential arising from the planet and the gravitational force of the disk, respectively. Here, $M_p$ is the planet mass, $r'\equiv\abs{\mathbf{r}-\mathbf{r}_p}$ is the distance to the planet and $\epsilon$ is a softening length used to avoid a singularity of the potential in
the vicinity of the planet. In 3D disks when the MHD is included, we use a softening length of $\epsilon=0.2H(r_p)$. 

The equation of motion for the planet can be written as
\begin{equation}
\begin{split}
    \frac{d^2\mathbf{r}_p}{dt^2}=-\frac{G(M_\star+M_p)}{r_p^3}\mathbf{r_p}+\mathbf{f}_{\mathrm{disk}}&\\-2\mathbf{\Omega}\times\mathbf{v}_p-\mathbf{\Omega}\times(\mathbf{\Omega}\times\mathbf{r}_p)
    \label{eq:planet_motion}
\end{split}
\end{equation}
with the acceleration due to the gas disk
\begin{equation}
    \mathbf{f}_\mathrm{disk}=-G\int_V\frac{\rho(\mathbf{r})(\mathbf{r}_p-\mathbf{r})}{(r'^2+\epsilon^2)^{3/2}}dV
\label{eq:disk_pot}
\end{equation}
where $\mathbf{v}_p$ is the planet velocity. In Eqs (\ref{eq:ind_pot}) and (\ref{eq:disk_pot}), the integrations are performed over the whole volume $V$ of our computational domain.

%%%%%%%%%%%%%%---------
\begin{figure}
\includegraphics[width=0.5\textwidth]{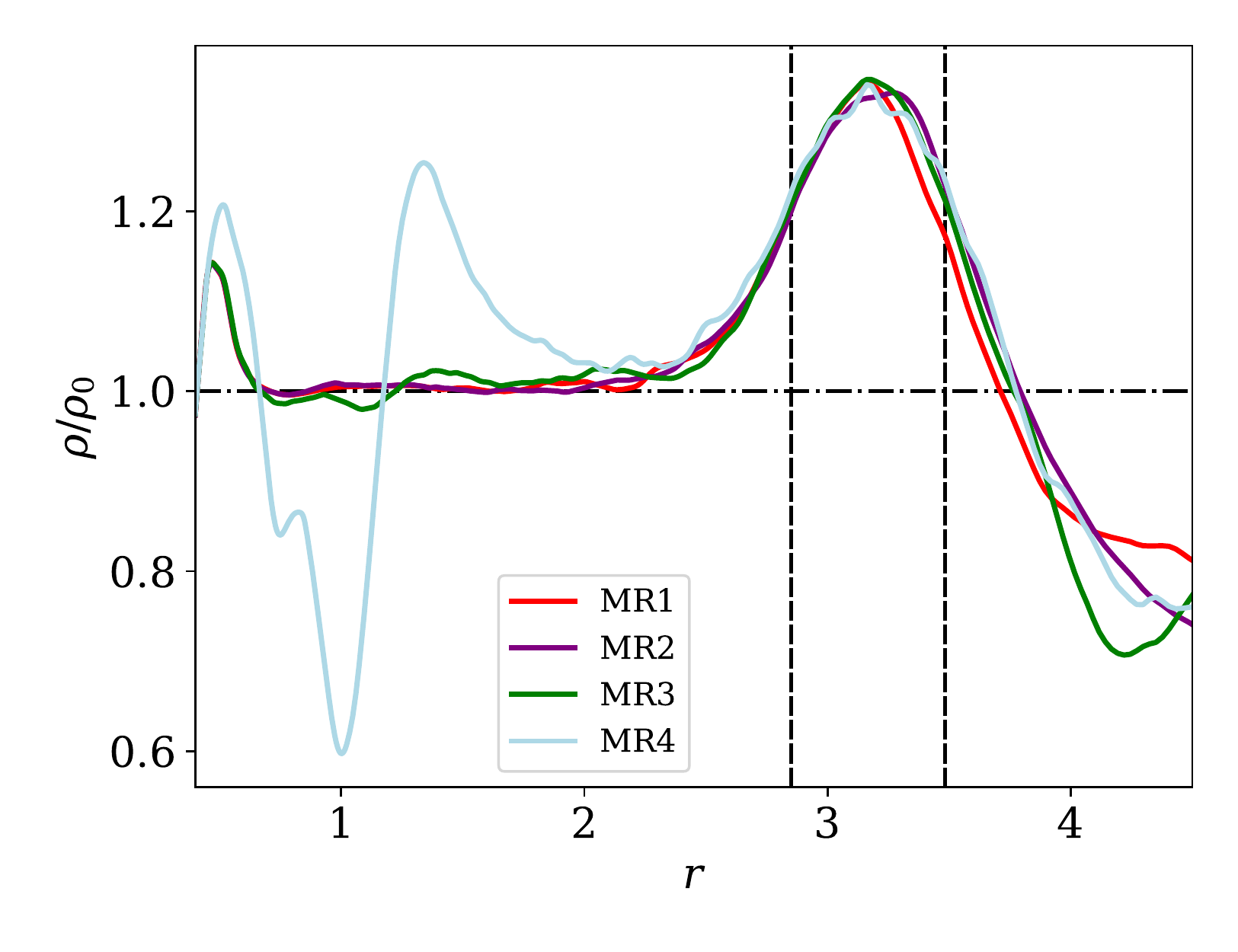}
\caption{Density bumps at $t=250\,T_0$ in our MHD simulations considering different planetary masses $M_p$ (MR1-MR4 models). Note the opening of a partial gap at $r=1$ and the formation of a secondary density bump due to the migrating planet with $M_p=100\,M_\oplus$. In this case, the dashed lines bracket $2H$ around the density maximum of the profile for $M_p=100\,M_\oplus$.}
\label{fig:dens_bumps}
\end{figure}
%%%%%%%%%%%%%%---------

%%%%%%%%%%%%%%---------
\begin{figure*}
\includegraphics[width=1.0\textwidth]{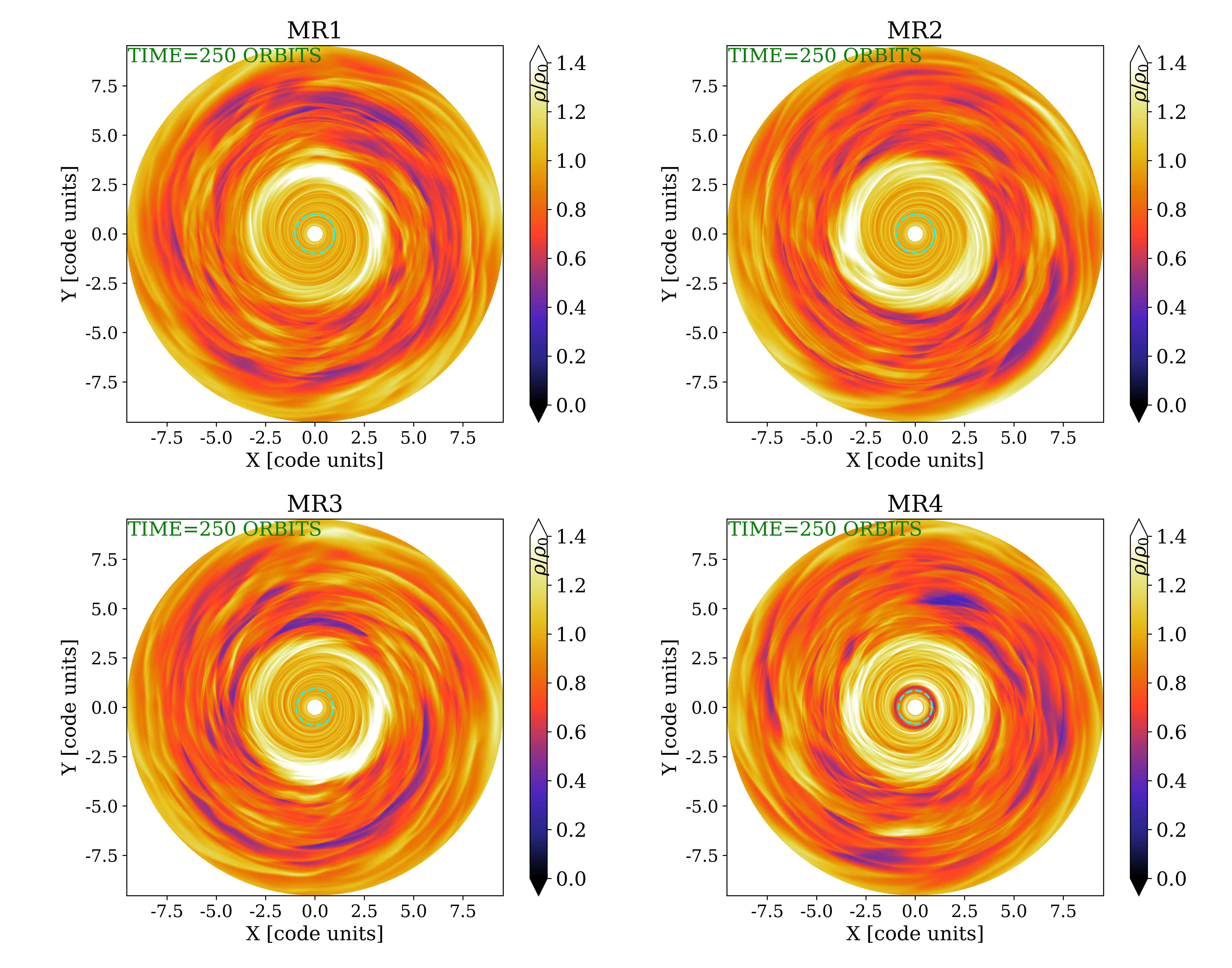}
\caption{Gas density in the midplane ($z=0$) of our MR1-MR4 models with resistivity transitions at $t=250\,T_{0}$.
Individual panels are labeled with the mass of the embedded planet $M_{\mathrm{p}}$. Vortex formation occurs at the 
transition between the dead and active zones of the disk and spiral waves appear propagating through the dead zone
where they interact with the migrating planet. The dotted green circles represent the planetary orbits.}
\label{fig:density}
\end{figure*}

In the case of our 2D $\alpha$-disks models, the continuity and Navier–Stokes equations are solved on a polar grid centred on to the star, and in a frame co-rotating with the planet. A locally isothermal equation of state is used with the vertically integrated gas pressure $P$ and the isothermal sound speed $c_s$ satisfying $P = \Sigma c_s^2$, with $c_s$ kept time-independent. The source terms in the Navier–Stokes equation include the gravitational potential due to the star, the planet, and the indirect terms arising from the stellar acceleration imparted by the disk gas and the planet (now the integrals in Eqs \ref{eq:ind_pot} and \ref{eq:disk_pot} are performed on a surface $S$
of the 2D computational domain). In the planet potential, a softening length $\epsilon= 0.6H(r_p)$ is included to mimic the effects of finite vertical thickness, with $H\equiv c_s/\Omega$ the disk’s pressure scale height. A viscous acceleration is also included with a turbulent kinematic viscosity $\nu$, which is related to the dimensionless alpha parameter by $\nu = \alpha c_s H$. It should be noted that the inclusion of this viscosity is motivated in principle by the results of the MHD simulations, since considering the resistivity transitions it is found that the Reynolds stress is not null within the deadzone \citep[see][]{Lyra_etal2015}. The initial radial profiles of the surface density and temperature are power-law functions with exponents $q_\rho$ and $q_T$, respectively. 

The main purpose of the presence of the 2D hydrodynamic simulations is only to compare the migration rate with our MHD models, therefore, details of planetary migration in viscous disks can be found elsewhere \citep[see for instance][and references therein]{Paardekooper_etal2022}. We recall that the width considered for the transition zone ($h_1=h_2=0.8$) does not produce any formation of vortices in the viscous simulation studied in \citet[][]{Lyra_etal2015}. Therefore, we have not included any viscosity transition zone for these models.

\subsection{Code and set-up} \label{sec:model}
To numerically solve Eqs
(\ref{eq:continuity}-\ref{eq:pressure}), we use the publicly available magnetohydrodynamic
code {FARGO3D\footnote{http://fargo.in2p3.fr}}
\citep[][]{Benitez_Masset2016} with MHD-orbital advection enabled \citep[see][for details]{Masset2000,Benitez_Masset2016}. The FARGO3D code solves the hydrodynamic equations with a time-explicit method, using operator splitting and upwind techniques on an Eulerian mesh. The update of the magnetic field governed by the induction equation (Eq. \ref{eq:induction}) is done by the method of characteristics \citep[MOC,][]{Stone_Norman1992} and the constrained transport method \citep[CT,][]{Evans_Hawley1988} is used to preserve the divergence-free property of the magnetic field. Orbital evolution of the planets is integrated by a fifth-order Runge-Kutta method (Cash $\&$ Karp 1990)
with a time step provided by the CFL condition of the MHD algorithm. 

Table~\ref{tab:condinit} summarizes the set of disk parameters and initial conditions used in our simulations. Although we consider the same extent of disk components as in \citet{Lyra_etal2015} where one has to $r=[0.4,9.6]r_0$, $\phi=[-\pi,\pi]$ and $z=[-0.1,0.1]$ and the same spacing (logarithmic in radius and uniform in azimuthal and vertical\footnote[1]{The vertical extent of the disk model used in this study has as its main objective to guarantee that the MRI is correctly resolved, and to be able to run the simulations at a longer orbital time at a reasonable computational cost.} directions, respectively), we use a higher resolution $(N_r,N_\phi,Nz)=(518,1024,64)$. We note that the results presented here are valid also for a lower resolution.

We use periodic boundary conditions in the azimuthal and vertical directions. To avoid reflections at the radial boundaries of our computational domain,
we use damping boundary conditions as in \citet{deVal2006}, the width
of the inner damping ring being $3.9\times 10^{-2}r_0$ and that of the
outer ring being $8.54\times 10^{-1}r_0$. The damping timescale at
the edge of each damping ring equals $1/20^{th}$ of the local orbital period. Following \citet{Lyra_etal2015}, we consider a net vertical magnetic field and ramp it smoothly to zero inward of $r=2.5$. The resistivity profile is given by Equation (\ref{eq:eta_pro}); the dead-to-active transition in the outer disk is located at $r_0=10$ au \citep[][]{Dzyurkevich_etal2013}, which in our scale free units translates to $r_0=1$. We set $r_1=2.5$, $r_2=10$ and $h_1=h_2=0.8$ in Eq. (\ref{eq:eta_pro}) in all our 3D-MHD simulations. Fig. (\ref{fig:eta_dens}) shows the resulting resistivity profile in which a smooth transition can clearly be seen in the range $r=1$--$4$, centered at $r=2.5$. The dashed vertical lines in Fig. (\ref{fig:eta_dens}) mark the radial region around the transition center $(\pm 2H)$ where the RWI occurs. 

%%%%%%%%%%%%%%%%%%%%
\begin{figure*}
\includegraphics[width=1.0\textwidth]{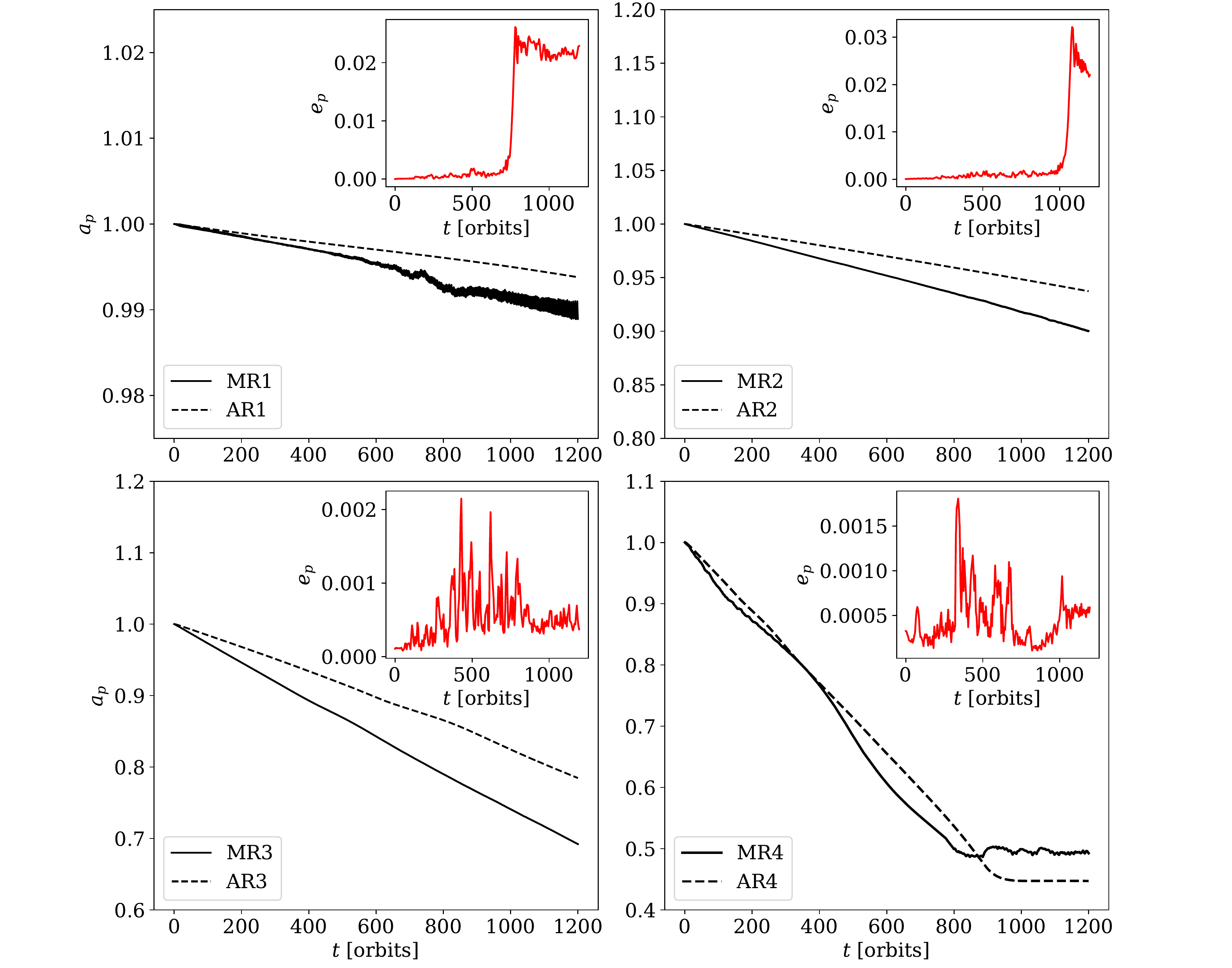}
\caption{Temporal evolution of the semimajor axis $a_p$ and eccentricity $e_p$ (inset box) of planets embedded in the dead zone of a disk where there are also spiral waves propagating towards the central star (for MR1-MR4 models). For comparison, we also show
the semimajor axes of planets migrating in 2D $\alpha$-disks (AR1-AR4 models).}
\label{fig:orb_parameters}
\end{figure*}
%%%%%%%%%%%%%%%%%%

\section{Results}
 \label{sec:results}
 
In this section we present the results of our 3D-MHD simulations considering a planet migrating in the dead zone of the protoplanetary disk. In addition, we investigate the gap opening by a massive planets in a fixed circular orbit. In order to demonstrate the differences between planetary migration affected by the background spiral waves with respect to migration in standard $\alpha$-disks, we also present 2D hydrodynamical viscous simulations that do not contain any type of transitions or additional perturbations in the dead zone (see Table \ref{tab:models} for a summary of the models). In these $\alpha$-disks, the value of kinematic viscosity $\nu_0$ is chosen in such a way that $\alpha\approx10^{-4}$ at the planet position. 

\subsection{Low-mass planet migration driven by background spiral waves} \label{subsec:low}

Fig. (\ref{fig:dens_bumps}) shows the radial density profiles as $t=250\,T_0$. Regardless of the planet
mass ($M_{p}=1$, $10$, $30$ and $100\,M_{\oplus}$), a density bump develops at the centre of the resistivity transition,
which gives rise to the RWI. For MR2, the density bump is slightly flattened. For MR4 (intermediate mass planet with $M_p=100\,M_\oplus$), a second peak appears at $r=1.4r_0$ due to the fact that the planet opens a partial gap.

In Fig. (\ref{fig:density}) we show the 2D map of the gas density in the midplane ($z=0$) at $t=250\,T_0$ of our MR1-MR4 models. In all cases, we see spiral waves excited by the turbulence in the active zone and by the RWI vortex at the resistivity transition. Consequently, the planetary migration can be modified by the background spiral waves because they generate a gas flow in the horseshoe region of the planet and additionally, they 
generate constructive (destructive) interference with the outer (inner) spiral waves emitted by the planet.

Fig. (\ref{fig:orb_parameters}) shows the evolution of the semimajor axis $a_p$ and the eccentricity $e_p$ for AR1-AR4 and MR1-MR4 models. We find that the migration of low-mass Earth-like
and Neptune-like planets ($M_p\leq30M_\oplus$) when MHD is included, is directed inwards and proceeds faster compared to the case of $\alpha$-disks.

%%%%%%%%%%%%%%%%%%%%%
\begin{figure}[ht!]
\includegraphics[width=0.5\textwidth]{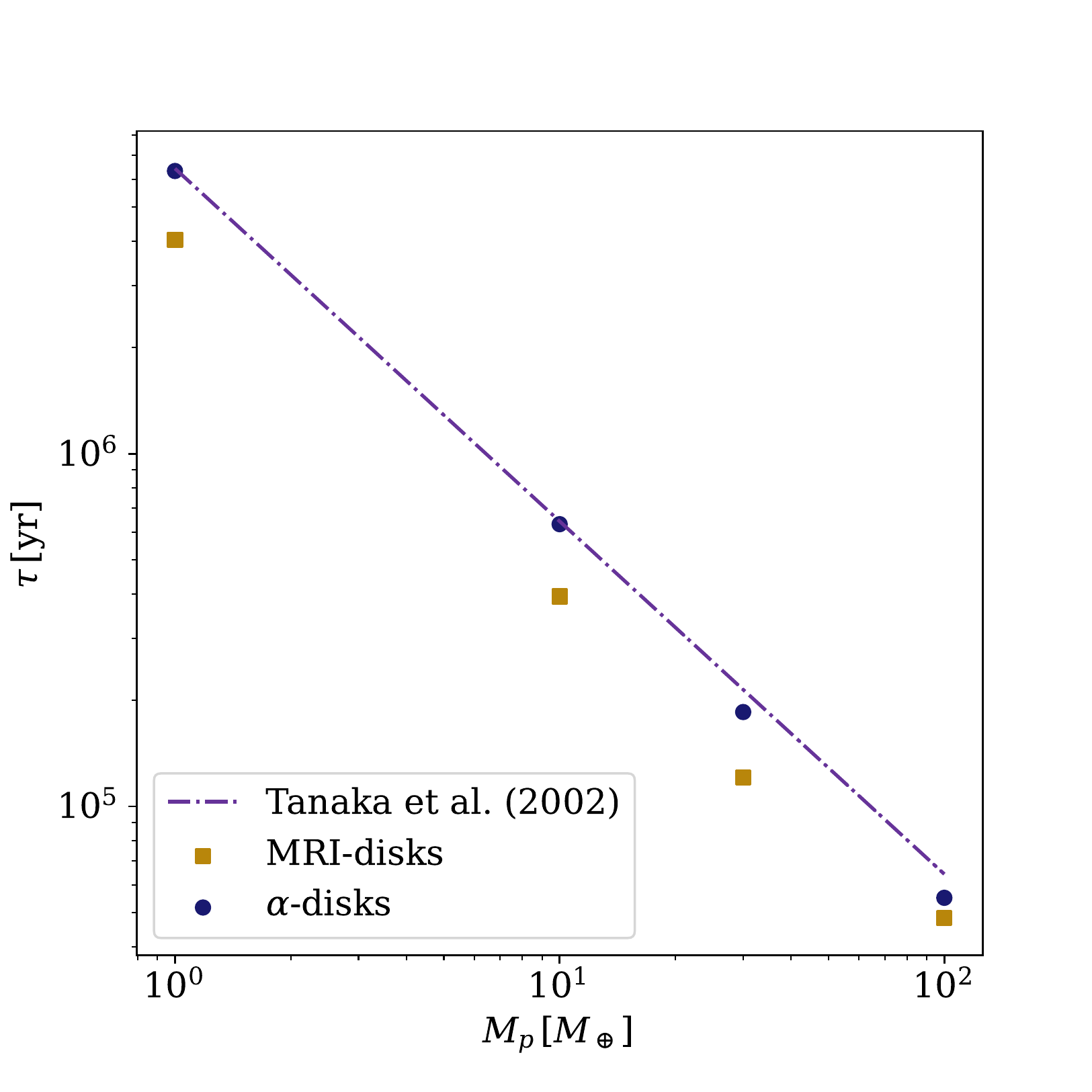}
\caption{Comparison of the migration time between the MRI-disks, 2D $\alpha$-disks (for AR1-AR4 and MR1-MR4 models) and the analytical results from \citet[][see Eq. \ref{eq:Tanaka}]{Tanaka_etal2002} as a function of the planet mass.}
\label{fig:tau}
\end{figure}
%%%%%%%%%%%%%%%%%%%%%

\subsubsection{Fast inward migration }

For a locally isothermal 3D disk, with a power-law surface density profile, the total torque (Lindblad plus corotation), $\Gamma$, acting on a planet was estimated by \citet{Tanaka_etal2002}. They found that
\begin{equation}
\Gamma=(1.364+0.541q_\rho)\Gamma_0,
    \label{eq:T3d}
\end{equation}
where $\Sigma_p$ and $c_s$ being the surface density and the sound speed at the planet position, respectively. Here, the total torque scales with $\Gamma_0\equiv(M_p/M_\star)^2(r_p\Omega_p/c_s)^2\Sigma_pr_p^4\Omega_p^2$. From equation (\ref{eq:T3d}), the type I
radial migration speed of the planet can be calculated from
the conservation of angular momentum, as

\begin{equation}
\frac{dr_p}{dt}=2r_p\frac{\Gamma}{L_p},
    \label{eq:rdot}
\end{equation}
where $L_p=M_p\sqrt{GM_\star r_p}$, is the planet's angular momentum. Using Equations (\ref{eq:T3d}) and (\ref{eq:rdot}), we can estimate the migration timescale as
\begin{equation}
 \tau=(2.7+1.1q_{\rho})^{-1}M_pr_p^2\Omega_p\Gamma_0^{-1}.
    \label{eq:Tanaka}
\end{equation}

To investigate the migration rates further, we compared the obtained values with $\tau=r_p(\frac{dr_p}{dt})^{-1}$ resulting from our simulations\footnote{The migration timescale was calculated between $t=0\,T_0$ and $t=600\,T_0$ in our numerical simulations, which seems to be a reasonable choice based on the behavior observed in the temporal evolution of the planetary semimajor axes (see Fig. \ref{fig:orb_parameters}).}, with the values given by equation (\ref{eq:Tanaka}).

As in 3D models of cylindrical turbulent disks \citep[see][]{NP2004}, we have been careful to consider a smoothing length for the planet potential appropriate for the 2D simulation set. This allows us to make a proper comparison with equation (\ref{eq:Tanaka}) for 3D isothermal disks and also gives us the possibility to make the comparison with our 3D-MHD models. The comparison of migration timescales is shown in Fig. (\ref{fig:tau}) as a function of $M_p$.
As expected, the 2D $\alpha$-disk simulations are in agreement with the migration time appropriate to 3D locally isothermal disks. In fact, the greatest difference ($14\%$) occurs between the AR4 and MR4 models.

Therefore, $\alpha$-disks as well as Eq.~(\ref{eq:Tanaka}) can provide reference values when quantifying how $\tau$ differs in the dead zone of disks when the MHD is included. Here, we find that the migration time for $M_p<100M_\oplus$ in $\alpha$-disks is $\approx 1.6$ times larger than the migration time of planets in disks where there are background spiral waves propagating through the dead zone. 

It is worth mentioning that the migration time of the planets does not depend on the strength of the initial magnetic field. We have analyzed the orbital evolution and the migration time for the MR2b model (those results are not shown here) and we have not found significant changes when the value of the initial magnetic field is increased.

In the case of the intermediate mass planet with $M_p=100M_\oplus$, we find that the migration is only slightly faster in the simulation when MHD is included for $t\leq800\,T_{0}$ ($\tau$ in the $\alpha$-disk simulation is $\approx$$1.15$ times larger compared to the 3D-MHD disk simulation). At $t>800\,T_{0}$, the migration becomes stagnant at a distance from the central star that is larger than in the $\alpha$-disk simulation (see panel 4 in Fig. \ref{fig:orb_parameters}).

\subsubsection{Time evolution of the eccentricity}

Furthermore, we find that when the MHD is included in our disk simulations, the eccentricity exhibits oscillatory behavior, and the amplitude
of oscillations varies with time. For MR1 and MR2, there are episodes when the eccentricity
suddenly becomes pumped up to $e_{p} = 0.02$--$0.03$ at $t=750T_0$ and $t=1050T_0$, respectively. 

%%%%%%%%%%%%%%%%%%%%
\begin{figure}[ht!]
\includegraphics[width=0.48\textwidth]{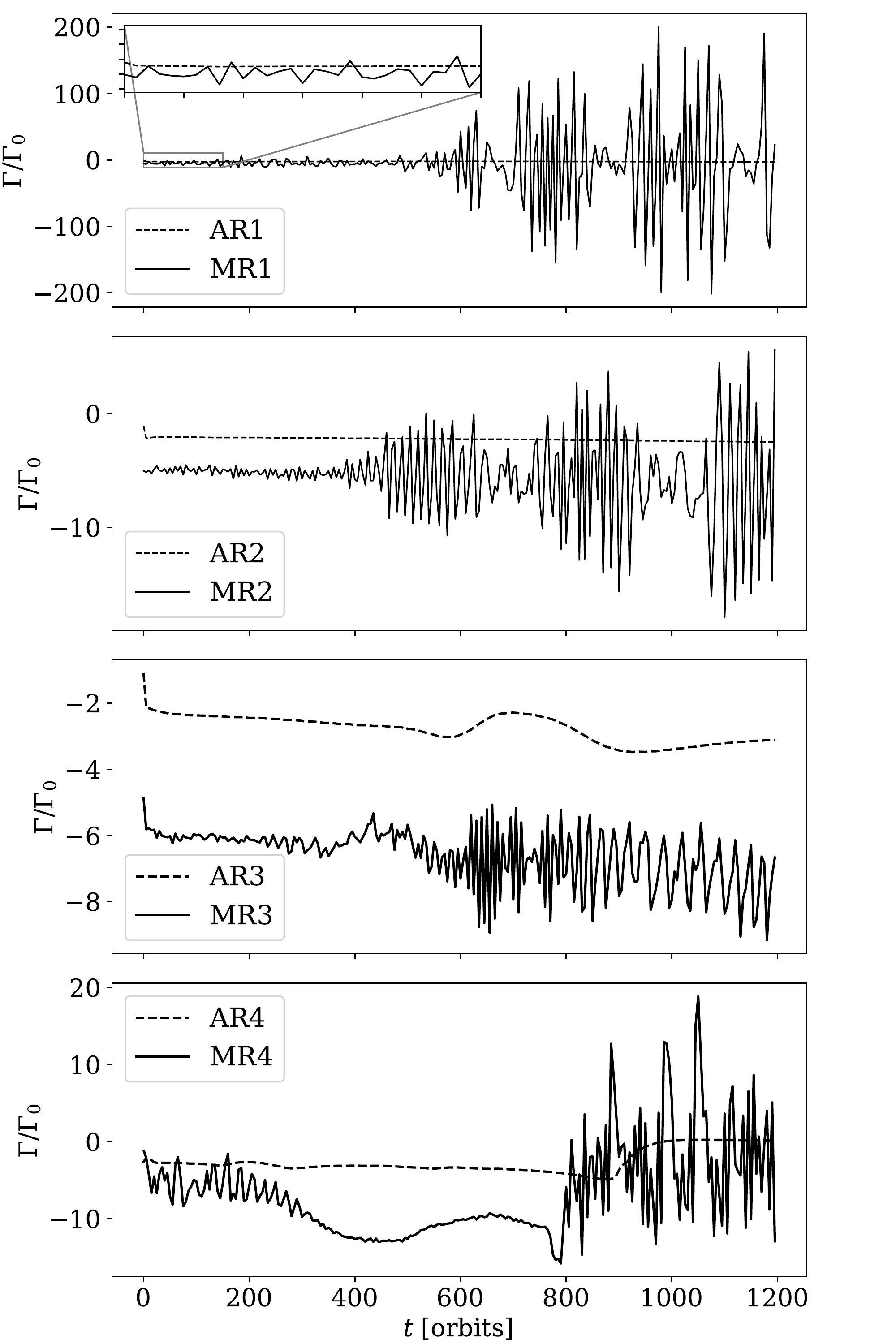}
\caption{Total torque versus time (in units of $\Gamma_0\equiv (\mu^2/h^2)\Sigma_pr_p^4\Omega_p^2$, here $\mu$ is the planet-to-star mass ratio), for AR1-AR4 and MR1-MR4 models. The oscillations are caused by the interaction of the spiral arms of the planet with the background spiral waves.}  \label{fig:torq}
\end{figure}
%%%%%%%%%%%%%%%%%%%%

This increase in eccentricity is due to transient vortices migrating from the outer edge of the dead zone inwards as they interact both directly and indirectly with the planet by modifying the background spiral waves. Once these vortices vanish, the eccentricity begins to decrease as can be seen in the upper right panel of Fig. (\ref{fig:orb_parameters}). For MR3 and MR4 models, the amplitude of oscillations can reach values around $10^{-3}$.

Although for planets embedded in $\alpha$-disks with masses greater than or equal to $20M_\oplus$ the damping time of the eccentricity differs a little from the linear theory, it is known that the eccentricity should drop exponentially to zero in a timescale $\tau_\mathrm{ecc}\propto(H/r)^2\tau$ \citep{Cresswell_etal2007}. However, we find that in our MHD-disk models the eccentricity of low-mass planets is not damped in the same way. In fact, it shows a slightly increasing behavior on average (at least for $M_p\leq30M_\oplus$, see inset boxes in Fig. \ref{fig:orb_parameters}).

%%%%%%%%%%%%%%%%%%%%%%%
\begin{figure}[t!]
\includegraphics[width=0.5\textwidth]{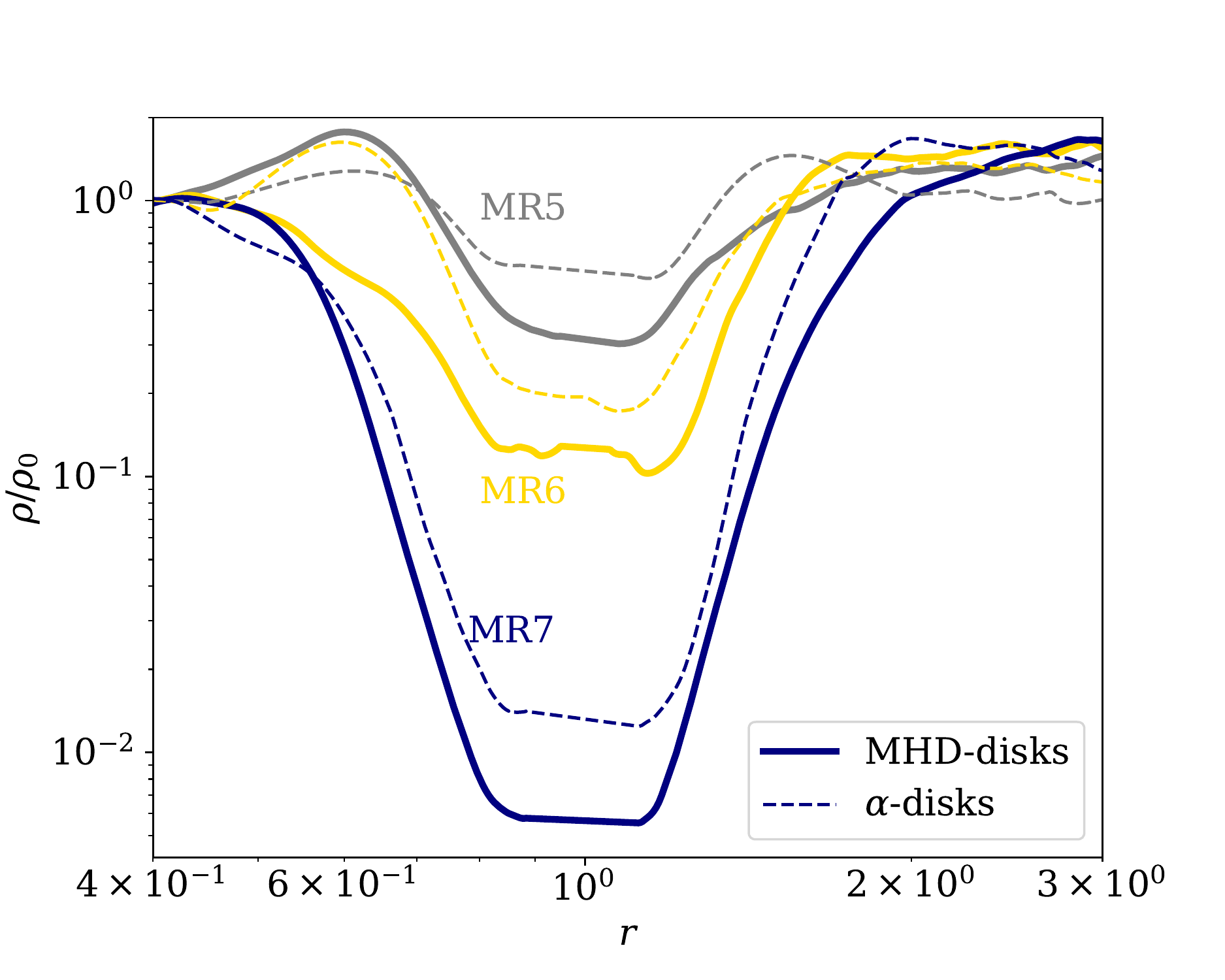}
\caption{Vertically and azimuthally averaged density profiles for 3D-MHD disks and azimuthally averaged density profiles for 2D $\alpha$-disks at $t=250\,T_0$.}
\label{fig:gap_profiles}
\end{figure}
%%%%%%%%%%%%%%%%%%%%%%%%

%%%%%%%%%%%%%%%%%
\begin{figure}[t!]
\includegraphics[scale=0.5]{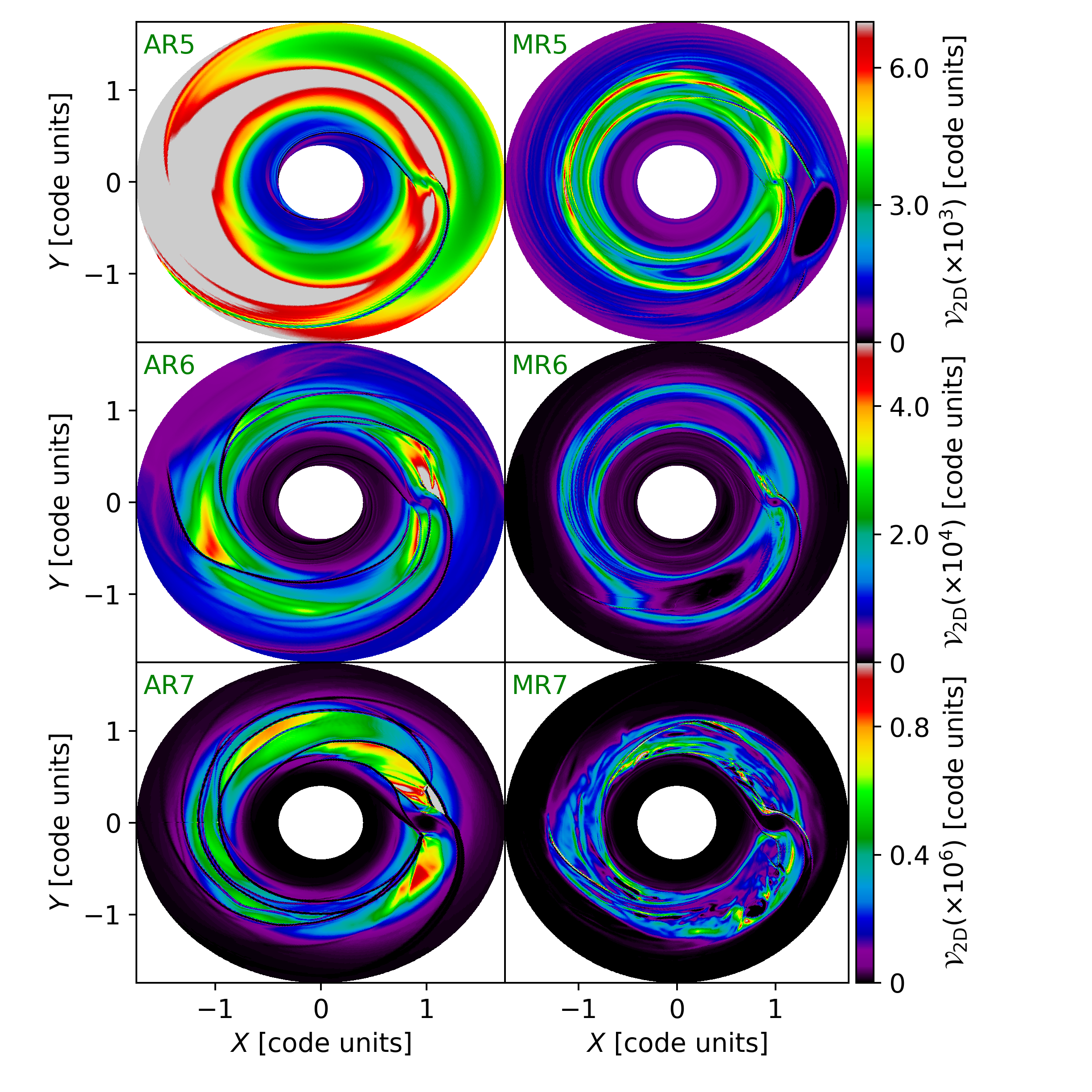}
\caption{Vortencity calculated at the midplane ($z=0$) for the AR5-AR7 and MR5-MR7 models at $t=250T_0$.}
\label{fig:gap_dens}
\end{figure}
%%%%%%%%%%%%%%%%%

%%%%%%%%%%%%%%%%%%%%
\begin{figure*}
\includegraphics[width=1.0\textwidth]{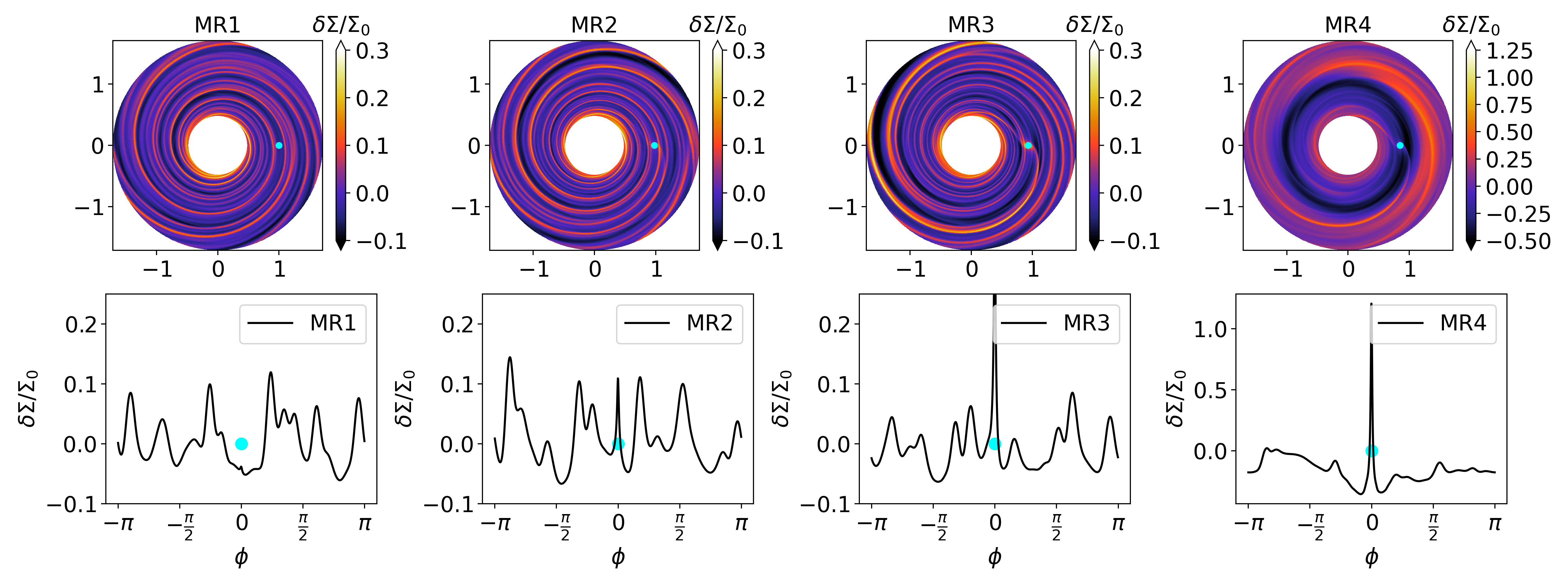}
\caption{\textit{Top row.} Perturbed gas density in the dead zone at the midplane ($z=0$) of the disk for the MR1-MR4 models, at $t=250T_0$. \textit{Bottom row.} Fractional amplitude of the spiral arms $\delta\Sigma/\Sigma_0$ at the radial planet position $r=r_p$ for these models. The cyan filled circle represents the position of the planet.}
\label{fig:deltaS}
\end{figure*}
%%%%%%%%%%%%%%%%%%

In addition, \citet{Nelson2005} found that the turbulence is the main source of eccentricity driving with a strong dependence on the planet mass (see his Figs. 5,6, 8 and 10). For instance, he found that a mass planet of $M_p=1M_\oplus$ can reach an eccentricity between $0.02$ and $0.08$ and for a planet with $M_p=10M_\oplus$ the value of the eccentricity oscillates in the range $0.02\leq e\leq0.03$, over the whole simulation time ($\sim500$ orbital periods). On the other hand, in this study we find that for $M_p\leq30M_\oplus$ the values of the eccentricities driven by the background spiral waves, although these also show an oscillatory behavior, are an order of magnitude smaller within the same time period than in \citet{Nelson2005}.

\subsection{Gap opening by massive planets} \label{sec:massive}
In this section, we study gap opening by massive planets in a fixed circular orbit, embedded in the dead zone. Since we are interested in analyzing the effect of background spiral waves on the shape and depth of the gap formed by a massive planet, we make the following considerations (in both, MHD and $\alpha$-disk models):
\begin{enumerate}
  \item[(i)] The planet is not allowed to migrate during the entire simulation time. One reason for this (apart from that related to viscous relaxation), is to see the evolution of the vortices that can form at the edges of the gap. 
  \item[(ii)] To avoid spurious shocks that would arise if the planet were initialised with its final mass, we introduce the mass of the planet smoothly as a function of time given by
  \begin{equation}
  M_p(t)= \left\{ \begin{array}{lcc}
             \frac{1}{2}\left[1-\cos{\left(\dfrac{\pi t}{t_\mathrm{ramp}}\right)}\right] &   \mathrm{if}  & t < t_\mathrm{ramp} \\
             \\ 1 &  & \mathrm{otherwise}. 
             \end{array}
   \right.
      \label{eq:Mp_t}
  \end{equation}
Here, we consider $t_\mathrm{ramp}=200T_0$, which is a suitable timescale for $\alpha$-disks \citep{Lega_etal2021}.
\end{enumerate}

With the above in mind, we consider the AR5-AR7 and MR5-MR7 models (see Table \ref{tab:models}). In Fig. (\ref{fig:gap_profiles}) we show the vertically and azimuthally averaged density profiles of the dead zone for disks when the MHD is included, as well as azimuthally averaged density profiles for the case of 2D $\alpha$-disks at $t=250\,T_0$. Focusing on AR5 and MR5 models first, a higher density bump is formed at $r=0.6$ in MR5. The reason is that at this orbital time, the gap depth is already considerable since the gap bottom has been emptied by a little more than $70\%$. On the other hand, at radii $r>r_p$, there is no well-defined density bump in the disk when the MHD is included compared to the $\alpha$-disk. Additionally, we find that the width of the gap is larger in the former case and we also notice that an increase in the density profile occurs from $r\approx1.8$ outwards. In summary, the planet in the dead zone when the MHD is included, is more efficient at removing gas from the gap region, that is, the density peaks at the edges of the gap are spread out and the bottom of the gap is deeper.

In fact, it can be seen in Fig. (\ref{fig:gap_profiles}) that the shape of the density profile for MR5 is somewhat similar to the density profile formed in AR6. In other words, a Jupiter-mass planet in the dead zone of a disk threaded by background density waves creates a gap with features that resemble a gap formed by a planet of 3 Jupiter masses in a disk governed by the classical viscous evolution. In the case of MR6 model, the density profile in the inner part ($r<r_p$), does not show any density bumps in comparison with the AR6 model. Similar to the previous case, for the outer part of the disk ($r>r_p$) in model MR6, the density profile shows an increase from $r\approx1.5$, however, the width and depth of the gap are greater than in AR6 model. 

On the other hand, for AR7 and MR7 models the density at the bottom of the gap it is very low. In this last case, we find that the shape of the gap in both models only differs in the outer part of the disk, since the gap in the $\alpha$-disk has a smaller width and shows a density bump at $r=2$.  Note that in the AR7 model there is again an increase in the density profile of the disk starting at $r\approx2.7$.

In Fig. (\ref{fig:gap_dens}) we show a comparison of the 2D-vortencity maps at the midplane $(z=0)$, between the AR5-AR7 and MR5-MR7 models at $t=250T_0$. Let us focus on the comparison of the MR5 and AR5 models. In the case of the AR5 model,  a large elongated vortex can be seen to form at the outer edge of the gap, while in the MR5 model in the same region a very weak vortex of much shorter length is observed (which may be the result of the formation of a transient vortex). We argue that at this orbital time ($t=250T_0$) the background spiral waves have removed enough density from the outer edge of the gap (and within the gap itself) which is accumulated on the inner edge of the gap, producing a density bump where the inner vortex forms. This vortex approaches its steady-state strength at $t=450T_0$ and survives until the end of the simulation time. Note that in both models there is formation of a vortex in the coorbital region behind the planet. Lastly, for more massive planets like those considered in the MR6-MR7 models, we did not find vortex formation at either of the two edges of the gap (see Fig. \ref{fig:gap_dens}), which may be due to flattening of the gap edges by the BSWs.

%%%%%%%%%%%%%%%%%%%%%%%
\begin{figure}[t!]
\includegraphics[width=0.5\textwidth]{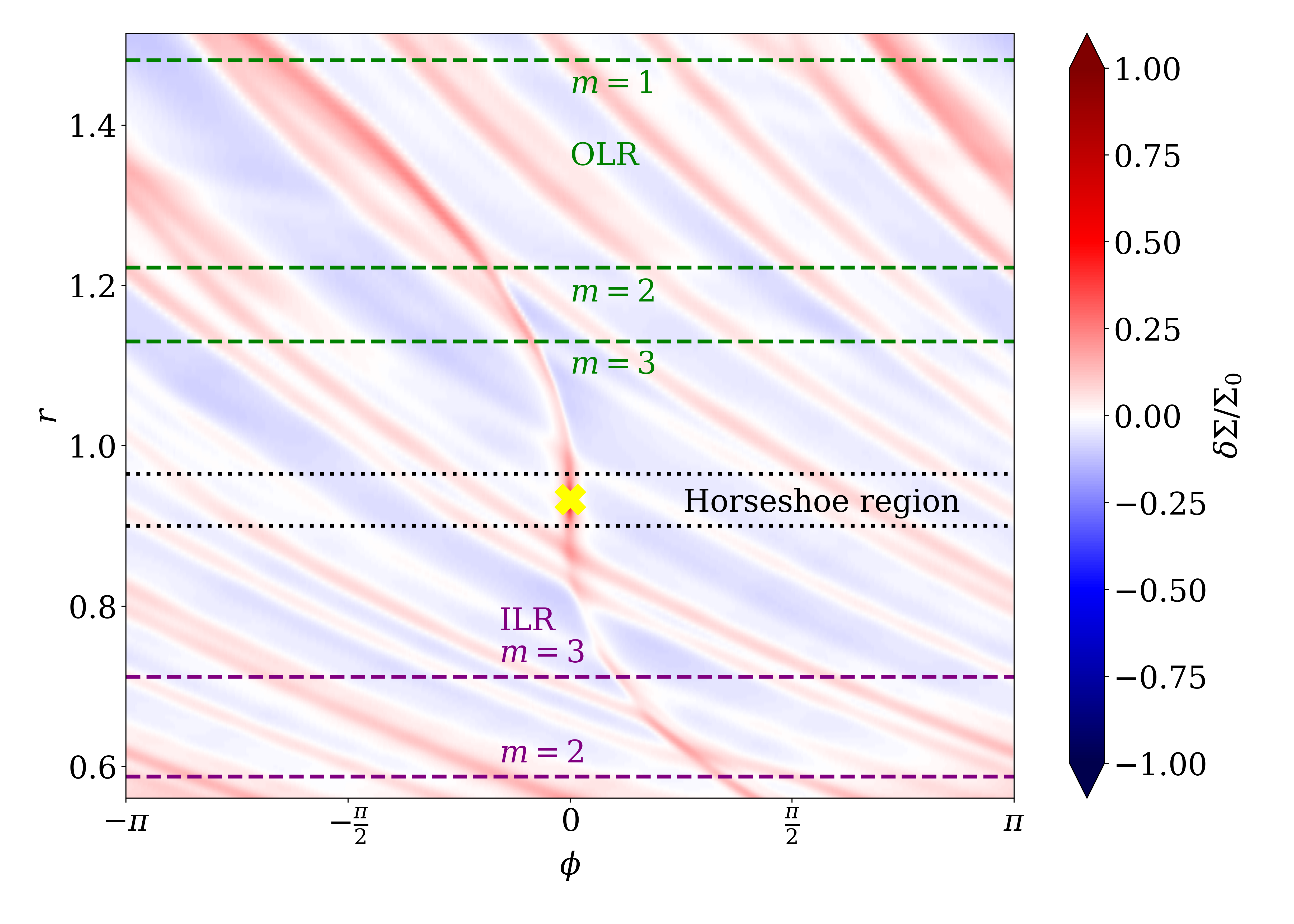}
\caption{Perturbed density in the dead zone of the disk for the MR3 model. The colored dashed lines shows various Lindbland resonance locations and black dotted lines show the horseshoe region and the yellow cross marker shows the planet's position.}
\label{fig:LR}
\end{figure}
%%%%%%%%%%%%%%%%%%%%%%%%

\section{Discussion} \label{sec:discussion}
From our results presented above it can be argued that the background spiral waves effect can substantially modify the migration of low to intermediate mass planets. We show that when there is a set of spiral waves propagating in the dead zone of the gas disk, they can accelerate the inward migration of low-mass planets. In this section, we investigate and discuss the background spiral waves effect on Lindbland and corotation torques, in order to understand the main cause of such fast inward migration.

\subsection{The background spiral waves interference effect}

Although the spiral waves generated by a vortex shows a spiral wave pattern which is reminiscent of an embedded planet, they do not have the same origin. The former, is the result of Rossby waves which are vorticity waves propagating on the gradient of vortensity \citep[][]{,Lovelace_etal1999,Li_etal2000,Meheut_etal2013}. From linear theory, their dispersion relation is given as \citep{Meheut_etal2013}
\begin{equation}
    \Tilde{\omega}=\frac{2\kappa^2c_s^2}{c_s^2(k_r^2+m^2/r^2)+\kappa^2}\frac{m\mathcal{L}'}{r\Sigma},
    \label{eq:dispersion}
\end{equation}
with a wave frequency $\omega$. In equation (\ref{eq:dispersion}), $\Tilde{\omega}\equiv\omega-m\Omega$ is the wave frequency in a rotating frame, $\kappa$ is the epicyclic frequency, $m$ and $k_r$ are the azimuthal and radial numbers, respectively, and $\mathcal{L}'$ represents the radial derivative of the quantity $\mathcal{L}$ (which is related to the inverse of vortencity). Due to differential rotation, the Rossby waves are coupled to the density waves \citep{Tagger2001,Bodo_etal2005} and the latter can only propagate outside of Lindblad resonance positions given by
\begin{equation}
    \omega-m\Omega(r_{LR})=\pm\kappa(r_{LR}).
    \label{eq:Lindbland_radius}
\end{equation}

In fact, when the linear phase of the Rossby waves saturates, the amplitude of the density spiral waves is higher at the Lindblad resonances closer to the corotation radius (defined for the vortex) \citep{Meheut_etal2013}.

Note that we are interested in the amplitude of these density waves when they interfere (constructively or destructively) with the spiral waves emitted by the planet. Therefore, a study to parameterize the amplitude and non-linear saturation criteria of background spiral waves (created by RWI and from turbulence in the active region of the disk) are beyond the scope of this work. With all of the above in mind, we now focus on discussing how is the background spiral waves interference with the (inner/outer) spiral arms of the planet.

In Fig. (\ref{fig:deltaS})  we show the shape of the background spiral waves in the dead zone of the disk (top row) and their physical amplitudes, $\delta\Sigma/\Sigma_0\equiv(\Sigma-\Sigma_0)/\Sigma_0$, at the planet position $r=r_p$ (bottom row) for the MR1-MR4 models. In the 2D density maps of this figure it can be seen that when the mass of the planet increases, the spiral waves of the planet emerge in the density contrast and suffer interference from the background spiral waves.

%%%%%%%%%%%%%%%%%%%%%%%
\begin{figure}[t!]
\includegraphics[width=0.48\textwidth]{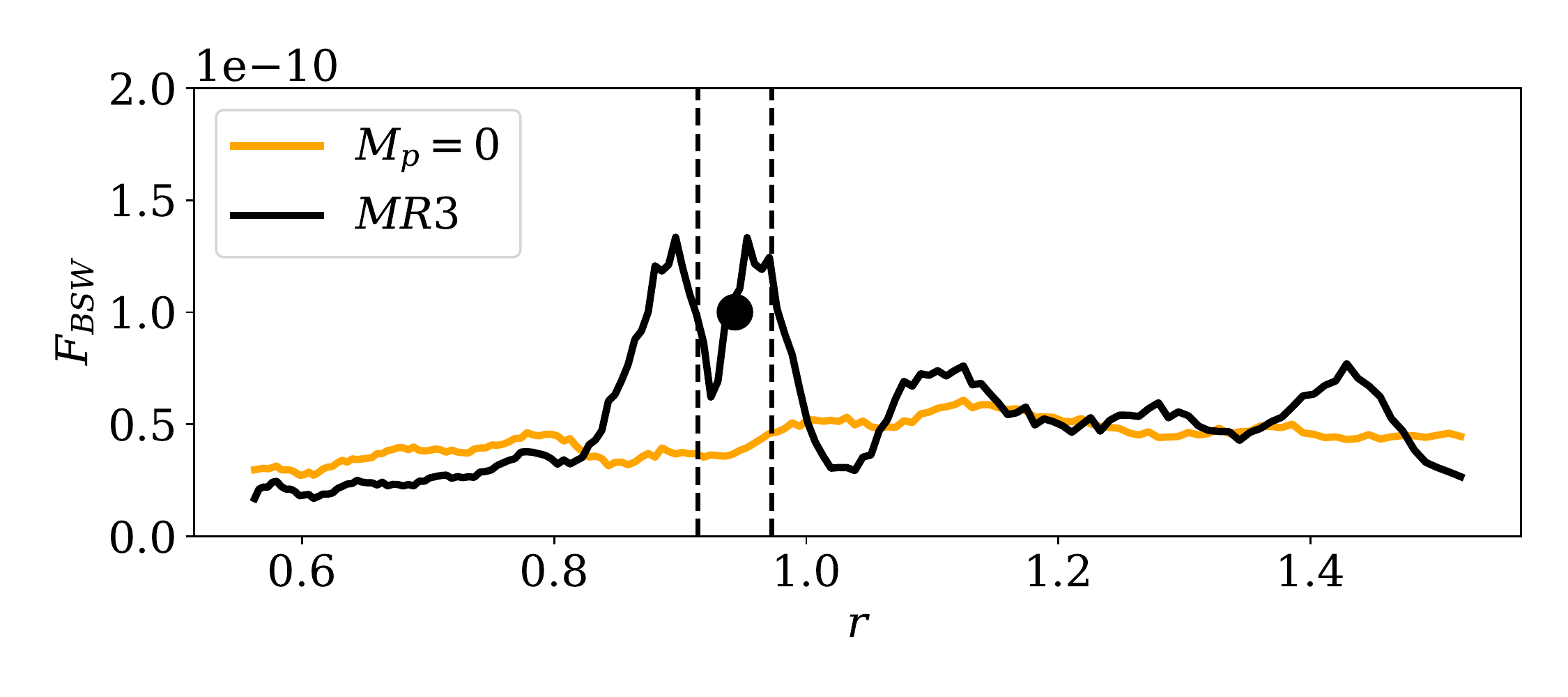}
\caption{Angular momentum flux through the dead zone of the disk for the MR3 model calculated at $t=210T_0$. The black dot and the black dotted lines represent the position of the planet and the width of the horseshoe region, respectively. The $M_p=0$ case is included for comparison purposes.}
\label{fig:ang}
\end{figure}
%%%%%%%%%%%%%%%%%%%%%%%%

Since the peak in the $\delta\Sigma/\Sigma_0$ curve located at $r=r_p$ and $\phi=0$ can be identified as the primary arm of the planet \citep{MR2019}, we can compare the amplitude of the background spiral waves with the spiral waves emitted by the planet for models MR1-MR4. From Fig. 9 (bottom row) it can be seen that for the MR1 model the amplitude of the background spiral waves is much greater than the density waves emitted by the planet, which explains why they cannot be seen in the 2D density maps. For the MR2 model both amplitudes are approximately equal. In the case of the MR3 model, the amplitude of the density waves emitted by the planet is greater than the amplitude of the background spiral waves, and the interference of the background spiral waves can be easily observed in the 2D map in Fig. \ref{fig:deltaS}. Lastly, for the MR4 model, the amplitude of the density waves generated by the planet is much greater than that of the background spiral waves, and because the planet begins to open a gap, the shape of the background spiral waves is modified, although its effect of interference remains. Note that as the amplitude of the waves emitted by the planet increases, the amplitude of the background spiral waves decreases.

Therefore, we conjecture that the fast migration found in 3D-MHD disks is mainly due to the fact that the background spiral waves interacts strongly with the spiral arms of the planet (and with the planet itself) substantially modifying the Lindblad's torque (and also corotation torque when is present, as we discuss in the next section). Specifically, these propagating toward the central star may eventually overlap with the planet's outer spiral arm (i.e.,
the pitch angle of the latter coincides with the pitch angle of the background spiral waves) and produce a more negative differential Lindblad torque. In other words, the outer spiral arm of the planet suffers from constructive wave interference.

As a demonstration of this hypothesis, let us consider the MR3 model. Fig. \ref{fig:LR} shows the background spiral waves and spiral waves of the planet. In this case, it can be clearly seen that the outer  planet's spiral arm suffers from constructive interference (since their pitch angles match), while the planet's inner spiral arm suffers from negative interference. In addition, it can be seen from this figure that the interference in the outer arm of the planet occurs at exactly some external Lindblad resonances, contrary to the inner arm, where the interference does not occur at the internal Lindblad resonances.

On the other hand, when the resonances are closer to the planet (ie., in the radial region $r_p\pm H$ which contribute more to the Lindblad torque), we find that while constructive interference from the outer arm of the planet with the background spiral waves do not make a total match, the latter produce a redistribution of density in this region that drives the strengthening of the outer wake of the planet continuously.

In other words, the effect of the background spiral waves builds the density towards the planet when $r>r_p$ and pushes it away from the planet for $r<r_p$. In addition to producing a greater flux of angular momentum in the outer part of the dead zone ($r>r_p$). To verify the latter, we have calculated the angular momentum flux in the dead zone of the disk for the MR3 model at $t=210T_0$ by means of the expression

\begin{equation}
F_{BSW}=\int_0^{2\pi}(\Sigma r v_\phi ')v_r' rd\phi
    \label{eq:FJ}
\end{equation}
where $v_r'=v_r-\Bar{v}_r$ and $v_\phi'=v_\phi-\Bar{v}_\phi$, with $\Bar{v}_r$ and $\Bar{v}_\phi$ the radial and azimuthal components of the velocity averaged azimuthally. The result of this calculation is shown in Fig. \ref{fig:ang} where we also included the case of the flux of angular momentum in the disk without a planet. A direct comparison of the angular momentum flux in the disk for the MR model with the case of $M_p=0$ (no planet in the disk) shows a larger angular momentum flux on the outer part of the dead zone when the interference effect of BSWs takes place.

Here it is important to emphasize that the angular momentum flux is carried by a set of spiral waves launched "continuously" by the vortex at the transition zone and the turbulence of the active zone. Therefore, the planet's spiral arms can interact many times with the same or different background spiral waves within a planet's orbital period, unlike an interaction of a planet with the spiral arms of a second planet for example, where the interaction occurs in the horseshoe region \citep[see][]{Cui_etal2021}.

%%%%%%%%%%%
\begin{figure}[ht!]
\includegraphics[width=0.5\textwidth]{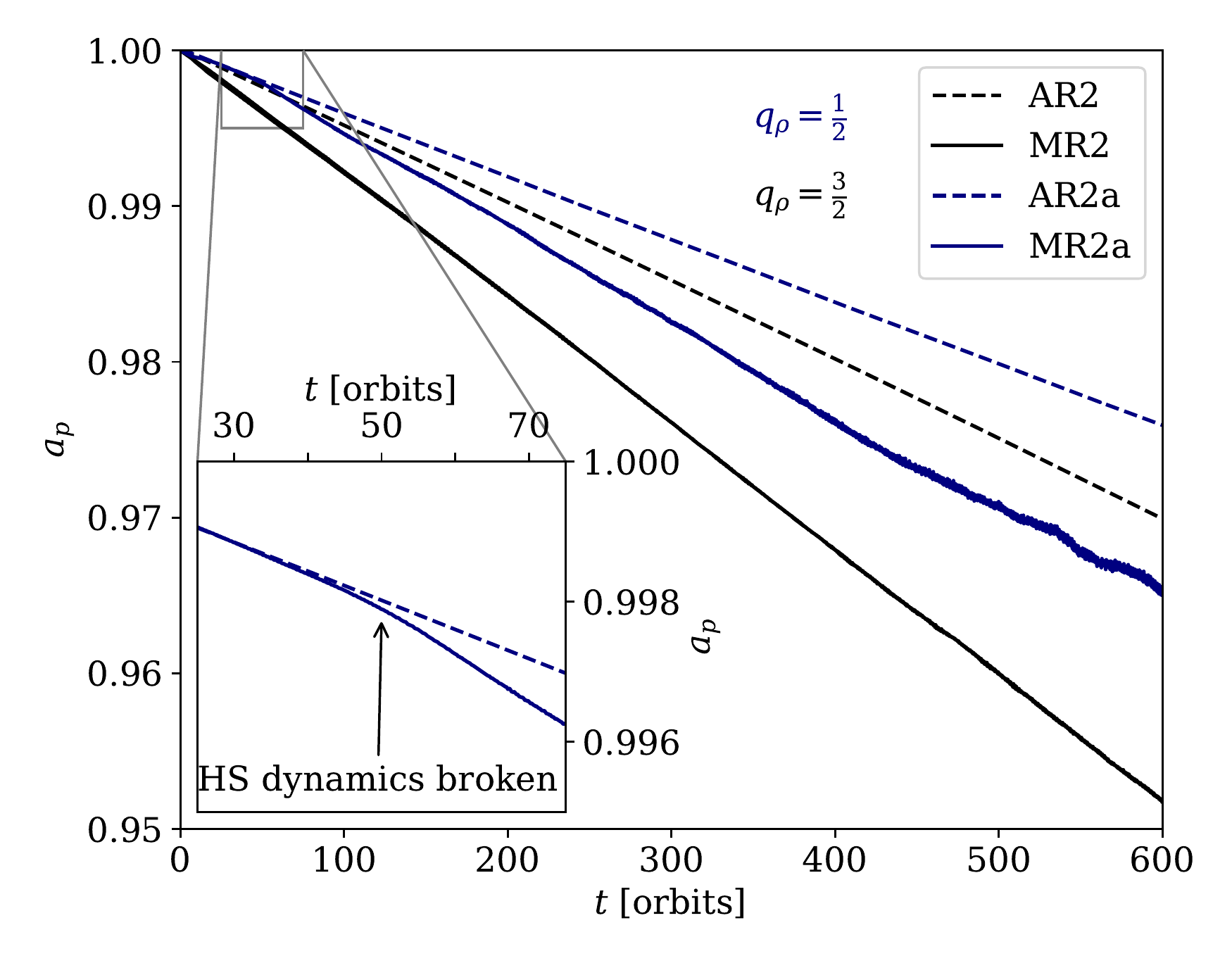}
\caption{Temporal evolution of the semimajor axis of the planet for AR2, AR2a, MR2 and MR2a. For both models of disks we use two different slopes for the radial density of the disk ($q_\rho=\{1/2,3/2\}$). In the zoomed region (\textit{lower left corner}) we show the behavior of the semimajor axis of the planet for the MR2a model when the dynamics of the horseshoe region (HS) is destroyed.}
\label{fig:ap_ss}
\end{figure}
%%%%%%%%%%%%

\subsection{A more negative oscillating Lindblad torque}

As we mentioned earlier in the introduction, several studies on turbulent disks have shown that the total torque on a migrating planet can present a strongly oscillatory behavior, leads to a random walk migration \citep{NP2004,PNS2004,LSA2004,Nelson2005}.

The main cause of random walk migration are the density fluctuations in the disk, in turn causes in most cases that the wakes generated by the planet are also turbulent. Furthermore, density perturbations generated in the turbulent zone of the layers in a disk can propagate into the dead zone and drastically change the torque on the planets \citep{OMM2007}, which again can be stochastic as in a fully turbulent disk. 

In our case, we show that although the torques on the planet present both negative and positive oscillations (see Fig. \ref{fig:torq}), on average they result in a negative torque in most cases, thus generating a faster convergent migration towards the central star (see Fig. \ref{fig:orb_parameters}). 

Recently, \citet[][]{McNally_etal2020} showed that the migration of low-mass planets in near-inviscid 3D-disks, accelerates the inward migration due to the buoyancy torques arising as a consequence of the disk response through vertical motions to the planet potential at buoyancy resonances which leads to evolution of the vortensity of the libration region.

For a slow radial accretion flow, planet migration would still be accelerated by the evolution of the corotation torque. Since the buoyancy response of the disk causes the vortensity of the coorbital disk material to evolve too quickly.

Although, \citet{McNally_etal2020} found that
the gas flow rate in the disc surface layers \citep[which corresponds to fast flow case studied in][]{McNally_etal2018}, has minimal impact on inward migration, a laminar radial gas flow in the midplane induced by Hall effect \citep[see][]{McNally_etal2017,McNally_etal2018} may be a way to counteract the influence of the buoyancy response of the disk on the torque of the corotation.

It should be mentioned that although in this study we consider the density perturbations generated from the active zone of the disk, it can be said that they are density perturbations which show a well-defined pattern that induces a different effect on the torques on the planet compared to the constant radial gas flow studied in \citet{McNally_etal2017,McNally_etal2018}, and bouyancy torques \citet{McNally_etal2020}. Note that latter effect is only present in 3D stratified disks. Therefore, a direct comparison with the background spiral wave effect presented here is not possible.

Since we find that the effect of the spiral waves propagating in the dead zone of the disk mainly modifies the Lindblad torque when the corotation torque is not active. Therefore, a natural question that arises to investigate is what happens when the corotation torque is also present on the disk. In the next subsection we redirect this question.

\begin{figure*}
  \centering
      \includegraphics[width=\linewidth]{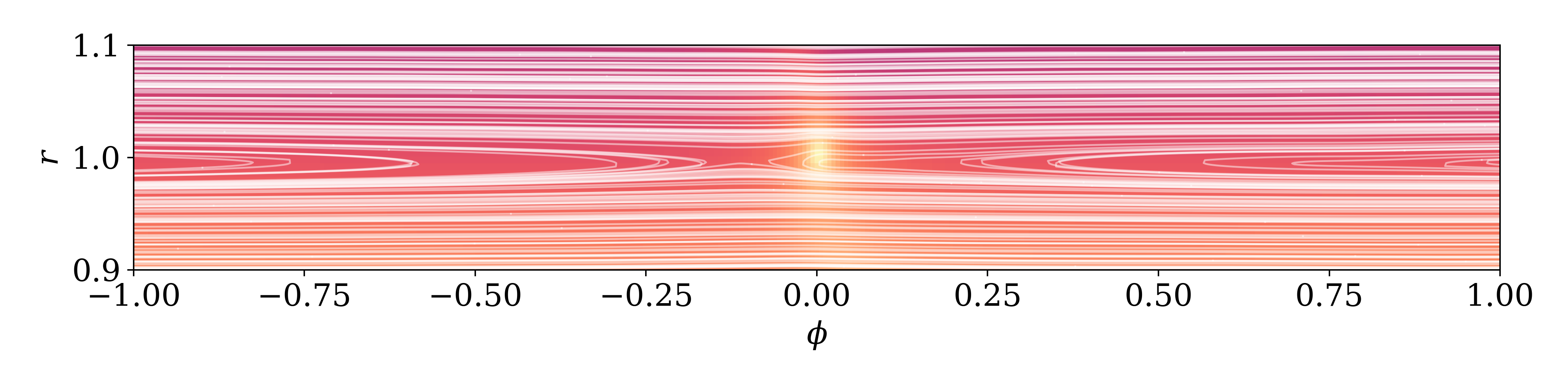}
      \includegraphics[width=\linewidth]{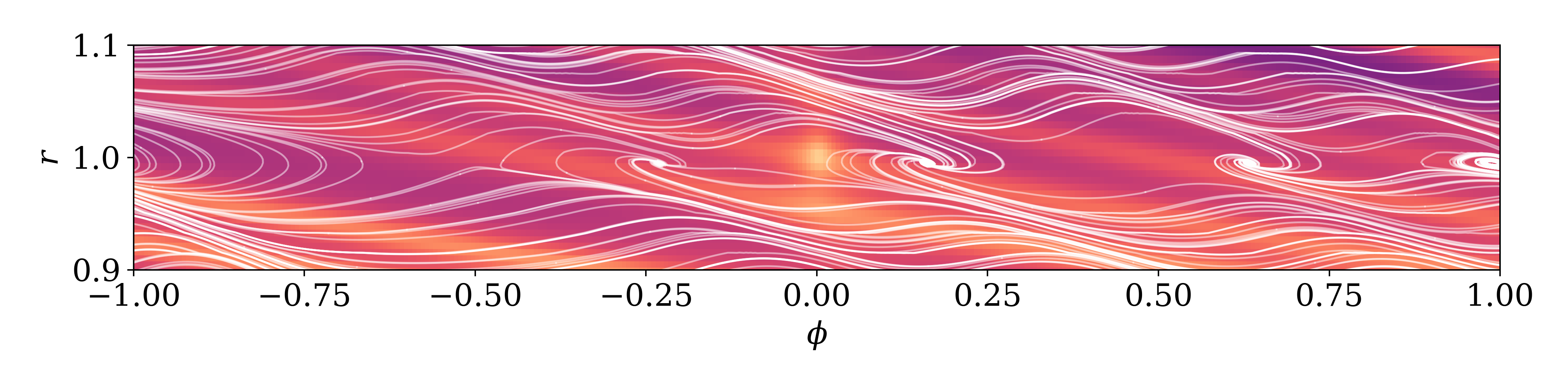}
      \includegraphics[width=\linewidth]{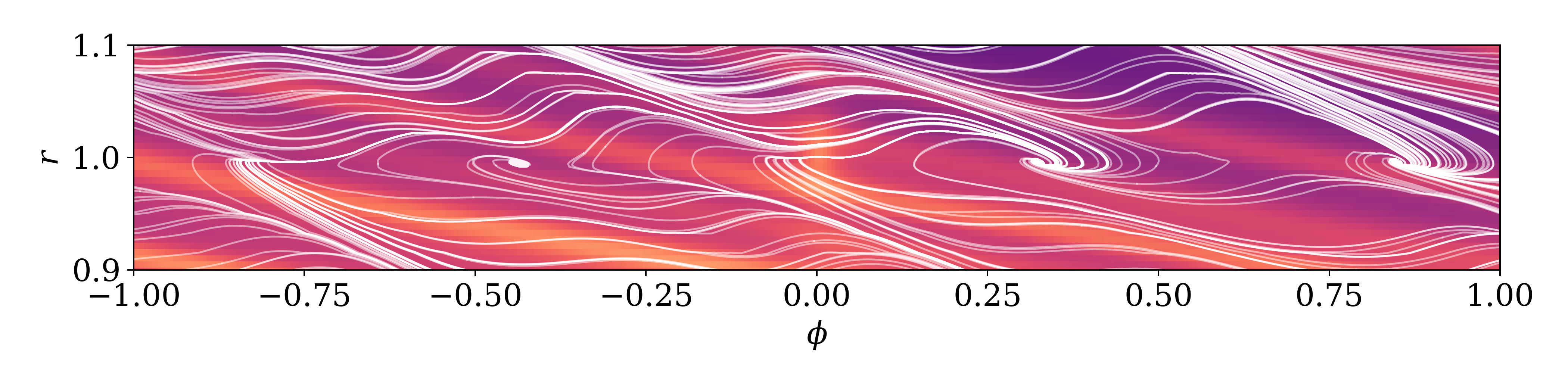}
  \caption{Streamlines overplotted in the gas density at the midplane for the MR2a model calculated at $t=5$, $t=60$ and $80$ orbits (top, middle and bottom panels, respectively).} 
  \label{fig:streamlines}
\end{figure*}

\vspace{5mm}
\subsubsection{Null vortensity gradient in the dead zone}

In the case of the corotation torque, it is well known that it can depend on the radial gradients of vortensity, entropy and temperature, as well as on the viscous production of vortensity \citep[][]{CM2009,MC2010,Paar_etal2010,BM2013,JM2017}. The latter arise from the contact discontinuities on the downstream separatrices if there is an entropy gradient \citep{JM2017}.

On the other hand, in an isothermal disk it holds that, if the surface density and angular frequency are power laws of the distance to the star, the vortensity gradient can become null for a particular value of the density slope \citep{JM2017}.

So, the choice of the value of the surface density slope $q_{\rho}=3/2$ (and that we consider a locally isothermal disk) gives us the possibility of being able to analyze in more detail the effect of the background spiral waves on the total torque on the planet, since the vortensity gradient (equal to $q_{\rho}-3/2$) is null.

Note that if the vortensity gradient is null, the contribution of the corotation torque to the total torque may be negligible, and therefore the total torque is governed only by the Lindblad torque. 

Finally, to support our hypothesis that the fast inward migration is mainly due to the effects produced by spiral waves propagating in the dead zone, and thus affecting the density in the horseshoe region and the spiral arms created by the planet, we performed a set of numerical simulations (MHD and $\alpha$-disks) considering a surface density of the disk which decays radially with $q_\rho=1/2$. 

In this way, the vortensity gradient is no longer zero and the corotation torque can make an additional contribution to the total torque which could modify the migration of the planet. Fig. (\ref{fig:ap_ss}) shows the temporal evolution of the semimajor axis of the planet for AR2 and MR2 models, in both cases with two different slopes for the radial density of the disk ($q_\rho=1/2$ and $q_\rho=3/2$).

\begin{figure*}
  \centering
    \includegraphics[width=\linewidth]{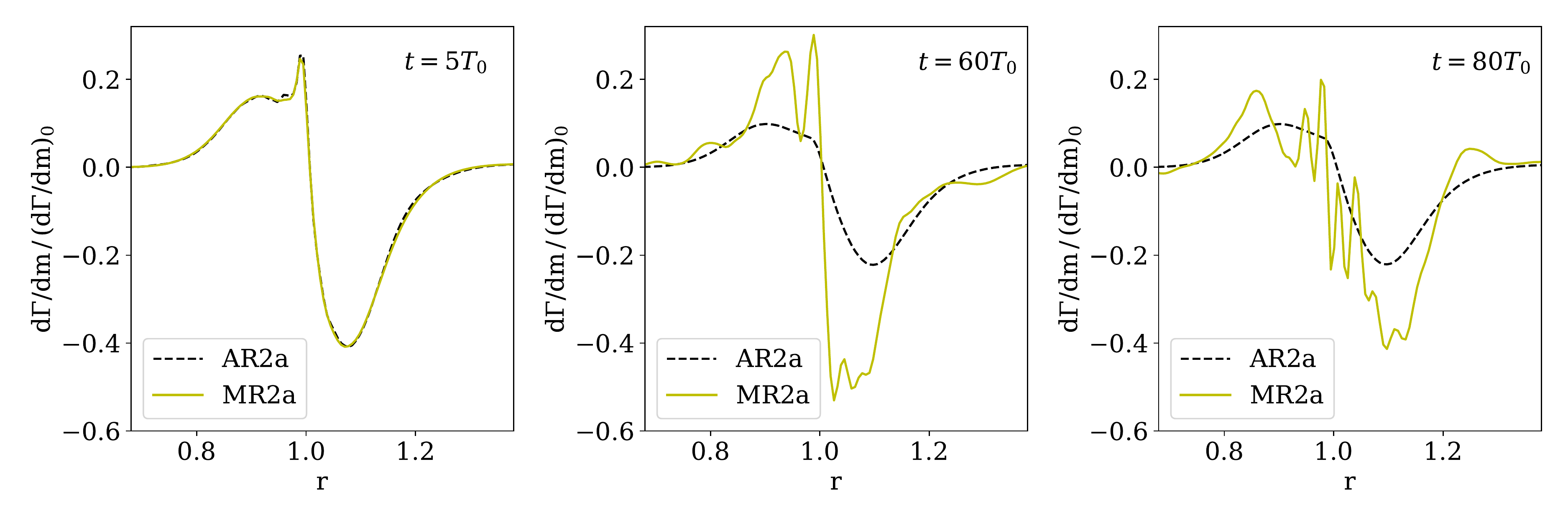}
  \caption{Radial torque distribution per unit disk mass in units of $(d\Gamma/dm)_0\equiv\Omega_p^2a_p^2\mu^2(H/a_p)^4$ for the AR2a and MR2a models at $t=5$, $t=60$ and $80$ orbits.} 
  \label{fig:dG_dm}
\end{figure*}

It is clear that when $q_\rho=1/2$ the corotation torque has a positive contribution to the total torque, so the migration of the planet is slower, but in both cases the relative difference in the migration most of the time it is very similar for both disk models ($\alpha$-disk and MHD-disks). However, particular attention should be paid to a time $t\lesssim50T_0$ in Fig. (\ref{fig:ap_ss}) for the MR2a model. Since within this time interval the migration is the same as for the AR2a model, this behavior can be explained by means of two effects on the corotation torque due to background spiral waves: 

\begin{enumerate}
  \item[(i)] The corotation torque still does not saturate and the perturbations of the background spiral waves are very weak. 
  \item[(ii)] This time interval is how long it takes for the background spiral waves to deactivate the corotation torque by destroying the dynamics of the horseshoe region. In other words, from $t>50T_0$ the dynamics of the horseshoe region no longer exists.
\end{enumerate}

We think that this behavior in the migration of the planets in both disk models is due to point (ii) (which should be studied in detail in a subsequent study, Chametla et al. in prep). Let us show a hint that the dynamics of the horseshoe region breaks down after a certain orbital period. In Fig. \ref{fig:streamlines} we show the streamlines overlaying the gas density at the midplane, calculated at $t=5$, $t=60$ and $t=80$ orbits. In a short time ($t=5T_0$) it can be seen that the streamlines define a clear pattern of the horseshoe region in the corotation region of the planet. At a time of $t=60T_0$ there is no well-defined behavior of the streamlines that form the horseshoe region, since much greater disturbance is seen from the BSW. Lastly, at $t=80T_0$ the streamlines do not define a horseshoe pattern.

To further support our statement that the fast migration found in the MR2a model with respect to the AR2a model is due to the destruction of the dynamics of the horseshoe region, we calculated the radial torque density. In Fig. \ref{fig:dG_dm} we show the results of this calculation for different orbital times. At $t=5T_0$ the torque is fully unsaturated since it shows a well-defined peak close to r=1, which is characteristic of the contribution of the corotation torque \citep[see][]{Kley_etal2012}. We find that behavior of the torque density at this short time in both models is basically the same. On the contrary, for orbital times of $t=60T_0$ and $t=80T_0$ we find that the torque density in the MR2a model shows a strongly oscillating behavior near of the planet. This oscillatory behavior of the torque density shows a lower positive contribution near the planet than in the case of the AR2a model. Note that for a time $t=80T_0$ the torque density for the AR2a model still remains in the unsaturated regime.

On the other hand, if it were due to the fact that the perturbations of the spiral waves are very weak, in the case when the corotation torque is not active ($q_\rho=3/2$) both planets should have a similar behavior at the beginning of their migration, which is ruled out because the semimajor axes are separate initially.

\subsection{The possible effects of the background spiral waves on the migration of massive planets}

The migration of massive planets embedded in gas disks is mainly governed by the opening of a gap. Recently, \citet{Lega_etal2021} studied the migration of massive planets in adiabatic gas disks considering a low viscosity. They found that the formation of a vortex at the outer end of the gap is feasible and conclude that two migration modes are possible, which differ from classical Type-II migration in the sense that they are not proportional to the disk’s viscosity. 

The first migration mode occurs when the gap opened by the planet is not very deep. This occurs when a big vortex forms at the outer edge of the planetary gap, diffusing material into the gap. The second migration mode occurs when the gap is deep so that the planet’s eccentricity grows to a value 0.2, which can produce a slower migration and even reverse it. With all this in mind, we have investigated the background spiral waves effect on the opening of the gap and the migration of a massive planet in the dead zone of an unstratified 3D-MHD disk with weak resistivity gradient. 

Our results show that the background spiral waves propagating inward can generate a greater flow of gas within the gap as well as induce the formation of a vortex at the inner edge of the gap for the MR5 model (see Fig. \ref{fig:gap_dens}), and for all the models studied (MR5-MR7) the background spiral waves prevent the formation of vortices at the outer edges of the gaps. Therefore, this could modify the migration of a massive planet with respect to the standard type-II migration and also regarding the migration found by \citet{Lega_etal2021}.

Remarkably, we find that the gap formed by a massive planet in the dead zone (where there are spiral waves propagating, and therefore a radial flow of gas towards the inner part of the disk), is asymmetric. 

Considering such asymmetry in the gap and the spiral waves passing through the gap could definitely unlock the planet from the viscous evolution of the disk to give rise to a different behavior in their migration. 

Although the fact that massive planets embedded in turbulent disks can open a wider and deeper gap than in the case of a laminar disk was already reported by \citet{NP2003} and also by \citet{Zhu_etal2013}, the novelty of our results is that a massive planet can also open a wider and deeper gap in the dead zone of the disk, in addition to the fact that said gap can have an asymmetric shape.

As we mentioned in the introduction, in \citet{McNally_etal2019} a “city map" diagram of the different mechanisms that can modify planetary migration was introduced. We think that the effect of the background spiral waves could be located in the last three blocks of the upper left corner of this figure. Since as we have discussed before, our results suggest that the background spiral waves effect can modify the redistribution of the gas density in the dead zone, and therefore, it can change the total torque on low-to intermediate-mass planets (see Fig. \ref{fig:torq}). For massive planets it can modify the gas flow through the gap formed by the planet and also induce/prevent the formation of vortices at the edges of the gap depending on the planet mass.
It is important to note that the effect of background spiral waves is in part the result of RWI in the inner/outer transition regions of the disk dead zone as several previous studies have shown \citep[][]{Gammie1996,VT2006,TD2009,Lyra_Mac2012,Lyra_etal2015,Faure_etal2016}. Therefore, it can be considered a natural and frequent process in the temporal evolution of the protoplanetary disk.

\subsection{Comparison with low-mass planets embedded in full turbulent disks}

\citet{Nelson2005} has studied the orbital evolution of low-mass planets ($M_p\leq30M_\oplus$) embedded in unstratified full turbulent, magnetized disks. Although we cannot make a direct comparison between our results and those presented in said study, since there are minor differences. This work allows us to show the main differences in the orbital evolution of a planet migrating in the dead zone of the disk, perturbed by background spiral waves, with respect to its orbital evolution in a full turbulent disk. \citet{Nelson2005} found in all his models that the planets undergo a stochastic migration. Whereas we find (at least for the same range of planet masses) a well-defined pattern of rapid migration towards the central star (see Fig. \ref{fig:orb_parameters}). 

Another marked difference between results presented in \citet{Nelson2005} and our results, is that the long term torque fluctuations dominate over the type I migration torques. In other words, the torque felt by a migrating planet in a turbulent disk is less negative (even positive) than in a viscous laminar disk (see his Fig. 12). While we find that although the torque on the planet when the MHD is included in the disk has an oscillatory behavior, said torque on average does not exceed the torque in an $\alpha$-disk (the latter is less negative), as can be seen in Fig. \ref{fig:torq}.

Note that we have not included in this comparison the case when the planet has a mass of $M_p=30M_\oplus$, since \citet{Nelson2005} includes 3 planets in the same simulation and although they do not get too close to each other, the turbulent density fluctuations are of similar amplitude to the spiral wakes generated by the planets in this case, and therefore, the spiral waves emitted by each of the three planets can also modify the total torque and cause a different migration \citep[][]{Baruteau_etal2014,Cui_etal2021,Chametla_Chrenko2022}.

\subsection{Caveats and future model improvements}
\label{subsec:weaknesses}

It is important to note that we have neglected the dependence of the gravitational potential on $z$ along with the vertical stratification of the disk, for reasons of computational cost. Thus our simulations are of cylindrical disks \citep[e.g.][]{Armi1998,H2000,H2001,SP2002}. These simulations have been used in previous studies to show the effects of the magnetic field on gas dynamics and for planet-disk interaction in global simulations \citep{NP2003,NP2004,Nelson2005,Lyra_etal2015}.

Although, these type of simulations have the disadvantage that can overestimate the results of the total torque on a planet (which applies to our models), they also give us the possibility to run our numerical models more larger orbital time. 

The question of how much the fast migration toward the star can change if the vertical stratification is included, must be redirected in a subsequent work (Chametla et al. in prep). However, it is reasonable to infer that the perturbations in density reported here can modify the total torque on the planet even if disk stratification and potential dependence on z are included, since the interference produced by spiral waves propagating in the disk on the spiral arms of the planet can produce an oscillating behavior of the torque even in 2D simulations \citep{Chametla_Chrenko2022}.

\section{Conclusions} \label{sec:conclusions}

%%%%%%%%%%%%%%%%%%%%%%%%%%
We have analyzed how background spiral waves propagating through the dead zones of protostellar disks affect the migration of low- to intermediate-mass planets and the opening of gaps in the disks' gas by embedded massive planets. The background waves arise at the transition radius separating the disk's dead zone from its magnetically-active region. Even shallow resistivity gradients here lead to strong gradients in the accretion stress \citep[][]{Lyra_etal2015}, producing a local steepening of the surface density profile that triggers the growth of the Rossby wave instability and the generation of anticyclonic vortices.

For migrating low-mass planets, the inward-propagating spiral waves speed the migration by up to a factor 1.6 over the rate in an $\alpha$-disk, while also exciting the planet's eccentricity.

For planets of intermediate mass, the spiral waves' effect on the total torque is less and the planet migrates only slightly faster than in an $\alpha$-disk.  The planet's migration however halts further from the central star, due to vortices excited at the edges of the partial gap opened by the planet.

For both low and intermediate masses, we find that the main torque on the planet is the Lindblad differential torque, since the planet's outer (inner) spiral wake is modified by the constructive (destructive) interference with the background spiral waves, because the background spiral waves “deactivate" the horseshoe region dynamics. However, our calculations do not rule out the possibility of an additional torque contribution from the horseshoe region in the presence of a vortensity gradient.

Finally, we find that the gap formed by a massive planet is shaped by the spiral waves passing through. The waves make the gap asymmetric and can excite (prevent) vortices in the inner (outer) gap edges. These effects together could either speed or slow the type-II migration of massive planets, depending on the rate of gas flow through the gap (governed by the background spiral waves) and the strength of the internal vortex. 
%\newline

%\begin{acknowledgments}
\hfill \break We are grateful to the referee for his constructive and careful report which significantly improved the quality of the manuscript. The work of R.O.C. was supported by the Czech Science Foundation (grant 21-23067M). The work of O.C. was supported by Charles University Research program (No. UNCE/SCI/023). Computational resources were available thanks to the Ministry of Education, Youth and Sports of the Czech Republic through the e-INFRA CZ (ID:90140). WL acknowledges support from NASA via grant 20-TCAN20-001 and NSF via grant AST-2007422. This work was carried out in part at the Jet Propulsion Laboratory, California Institute of Technology, under contract with NASA and with the support of NASA's Exoplanets Research Program through grant 18-2XRP18\_2-0059.
%\end{acknowledgments}

%%%\vspace{5mm}
%\facilities{HST(STIS), Swift(XRT and UVOT), AAVSO, CTIO:1.3m,


%\software{astropy \citep{2013A&A...558A..33A,2018AJ....156..123A},  
%          Cloudy \citep{2013RMxAA..49..137F}, 
%          Source Extractor \citep{1996A&AS..117..393B}
%          }



%-\appendix

%\section{Appendix information}


\bibliography{Manuscript}{}
\bibliographystyle{aasjournal}



\end{document}